\RequirePackage{ifpdf}
\documentclass[hyper,letterpaper]{JHEP3}
\usepackage{amssymb,amsfonts,bm,amsmath,empheq}
\usepackage{cite}
\usepackage{graphicx}
\usepackage{multirow}
\usepackage{verbatim}
\usepackage{appendix}

\usepackage{url}
\usepackage{float}
\usepackage[T1]{fontenc}

\newcommand{\bea}{\begin{eqnarray}}
\newcommand{\eea}{\end{eqnarray}}
\newcommand{\be}{\begin{equation}}
\newcommand{\ee}{\end{equation}}

\def\IZ{\mathbb {Z}}

\def\IR{\mathbb {R}}

\def\IP{\mathbb {P}}

\renewcommand{\hat}{\widehat}

\title{Singular vector structure of quantum curves}

\author{Pawe{\l} Ciosmak$^{1}$, Leszek Hadasz$^{2}$, Masahide Manabe$^{3}$ and Piotr Su{\l}kowski$^{3,4}$\\
$^1$ Faculty of Mathematics, Informatics and Mechanics, University of Warsaw, ul. Banacha 2, 02-097 Warsaw, Poland  \\
$^2$ M.\ Smoluchowski Institute of Physics, Jagiellonian University, ul. {\L}ojasiewicza 11, 30-348 Krak{\'o}w, Poland  \\
$^3$ Faculty of Physics, University of Warsaw, ul. Pasteura 5, 02-093 Warsaw, Poland  \\
$^4$ Walter Burke Institute for Theoretical Physics, California Institute of Technology, Pasadena, CA 91125, USA}

\abstract{We show that quantum curves arise in infinite families and have the structure of singular vectors of a relevant symmetry algebra. We analyze in detail the case of the hermitian one-matrix model with the underlying Virasoro algebra, and the super-eigenvalue model with the underlying super-Virasoro algebra. In the Virasoro case we relate singular vector structure of quantum curves to the topological recursion, and in the super-Virasoro case we introduce the notion of super-quantum curves. We also discuss the double quantum structure of the quantum curves and analyze specific examples of Gaussian and multi-Penner models.
\\
\\
\\
\\
\\
\\
\\
\\
\\
{\tt CALT-2017-27}}

\begin{document}


\section{Introduction}

Riemann surfaces and plane algebraic curves, which can be defined simply by the equation
\be
A(x,y) = 0,     \label{Axy}
\ee
with $A(x,y)$ being a polynomial in complex variables, play an important role in modern theoretical and mathematical physics. They arise as spectral curves in matrix models, Seiberg-Witten curves, mirror curves in topological string theory, A-polynomials in knot theory, in various problems in enumerative geometry, and in many other contexts. In all these systems there is a natural notion of quantization, and algebraic curves mentioned above encode some information that can be interpreted as classical, which arises in the vanishing limit of the appropriate Planck constant, denoted $\hbar$ or $g_s$. This suggests that there exist quantum counterparts of those curves, referred to simply as quantum curves, which take form of differential (or difference) operators $\widehat{A}(\hat x, \hat y)$ annihilating appropriate partition functions $\psi(x)$
\be
\widehat{A}(\hat x, \hat y)\psi(x) = 0,      \label{Ahatxy}
\ee
with $\hat x$ and $\hat y$ satisfying the commutation relation
\be
[\hat  y, \hat x] = g_s.
\ee
In the classical, $g_s\to 0$ limit $\widehat{A}(\hat x, \hat y)$ is expected to reduce to $A(x,y)$. Such quantum curves and partition functions are also referred to, respectively, as Schr{\"o}dinger operators and wave-functions.

Examples of quantum curves have been found from several seemingly unrelated perspectives. First, they were identified in the system of intersecting branes, which encode Seiberg-Witten theory and provide quantization of Seiberg-Witten curves \cite{DHSV,DHS}. Second, it was realized that partition functions of branes in topological string theory satisfy difference equations, which quantize mirror curves \cite{ADKMV}. Third, it is known that determinant expectation values in matrix models satisfy differential equations, which can be regarded as quantizations of spectral curves \cite{abmodel,ACDKV}. Furthermore, analogous statements were found in knot theory in the context of the AJ conjecture, where colored knot polynomials turn out to satisfy difference equations that quantize A-polynomials \cite{Garoufalidis,GL}. Finally, enumerative geometry provides yet another source of such examples \cite{Norbury-quantum,Dumitrescu:2015mpa}.

While the above examples may seem unrelated, it appears that there is a quantization framework that works universally for all of them \cite{abmodel}, which is based on the topological recursion \cite{eyn-or}. The topological recursion was originally found as the solution to loop equations in matrix models \cite{Eynard:2004mh,Chekhov:2005rr,Chekhov:2006rq}, so it is not that surprising that it provides a quantization of matrix model spectral curves. It is however more surprising that the topological recursion provides a solution to the topological string theory on toric Calabi-Yau threefolds \cite{BKMP,Eynard:2010dh,vb-ps} -- and so, via geometric engineering, to Seiberg-Witten theory -- as well as reconstructs the asymptotic expansion of colored knot polynomials \cite{DijkgraafFuji-1,DijkgraafFuji-2,Borot:2012cw}. Based on these observations, a universal quantization procedure based on the topological recursion was proposed for all those and some other systems, with its universality explained by the underlying B-model structure (manifest possibly after some chains of dualities) of all those systems  \cite{abmodel}.

In the systems mentioned above, it is usually claimed that there is a unique quantum curve assigned to a given classical curve. However, as we summarize in this note following \cite{Manabe:2015kbj,Ciosmak:2016wpx}, it turns out that to a given classical curve one can naturally assign not just a single quantum curve, but an infinite family of quantum curves, which are in one-to-one correspondence with singular vectors (also known as null vectors) of the underlying symmetry algebra. In case of the original formulation of the topological recursion, and in all systems mentioned above (topological strings, Seiberg-Witten theory, knot polynomials, etc.) this is the Virasoro algebra. Quantum curves mentioned above and discussed so far in literature, from the perspective of \cite{Manabe:2015kbj}, correspond to Virasoro singular vectors at level 2. The relation of (previously considered) quantum curves to singular vectors at level 2 was noticed also in \cite{ACDKV}. However, in addition to those quantum curves at level 2 we identify an infinite family of quantum curves, which correspond to, and have the structure of, Virasoro singular vectors at higher levels. The first aim of this note is to discuss the structure and properties of such quantum curves, corresponding to all Virasoro singular vectors.

The second aim of this note is to present the relation between quantum curves and underlying symmetry algebra in another example, that of super-Virasoro algebra \cite{Ciosmak:2016wpx}. On one hand, this example nicely illustrates the general idea of the correspondence between quantum curves and symmetries. On the other hand, it leads to the construction of supersymmetric quantum curves, or super-quantum curves for short, which quantize underlying supersymmetric algebraic curves, and are interesting in their own right.

As mentioned above, the quantization procedure for algebraic curves can be very generally formulated by means of the topological recursion, at least in the context of the Virasoro algebra. However, to identify quantum curves corresponding to higher level singular vectors, it is of advantage to consider matrix model formulation of this problem, which is equivalent (at least once both matrix model and corresponding topological recursion formulations exist). In the matrix model formulation the relation to the underlying symmetry algebra is more explicit, and can be identified upon rewriting of the loop equations in the form of Virasoro-like constraints  \cite{Ambjorn:1990ji,Fukuma:1990jw,Itoyama:1991hz,Dijkgraaf:1990rs}. The approach based on matrix models and loop equations has also other advantages. First, it is not difficult to generalize this approach to the $\beta$-deformed case, which makes contact with conformal field theory with arbitrary central charge \cite{Awata:1994xd,Manabe:2015kbj}. Second, it enables to analyze other types of matrix models, with different underlying algebras. We will illustrate this statement in the example of a supersymmetric version of one-matrix model, the so called super-eigenvalue model introduced in \cite{AlvarezGaume:1991jd} and further analyzed in \cite{Becker:1992rk,McArthur:1993hw,Plefka:1996tt,Semenoff:1996vm,Itoyama:2003mv}, characterized by the underlying super-Virasoro algebra mentioned above. The topological recursion formalism is not known in this case, so it is of particular advantage to consider matrix model formulation. On the other hand, in the Virasoro case, having conducted matrix model analysis we reformulate the construction of higher level quantum curves in the language of the topological recursion, which is well understood in this case.

More precisely, our strategy is as follows. To start with, we identify a putative wave-function with some particular matrix integral, that we refer to as the $\alpha/\beta$-deformed matrix integral, or $\alpha/\beta$-deformed matrix model. It takes form of the $\beta$-deformed expectation value of an $x$-dependent determinantal expression, raised to a power parameterized by a parameter $\alpha$. We find that the wave-function defined in this way satisfies a differential equation in $x$ only for discrete series of values of $\alpha$, which turn out to coincide with degenerate momenta in conformal field theory interpretation. In case of the Virasoro algebra these values take form $\alpha=\alpha_{r,s}=-\frac{1}{2}(r-1)\epsilon_1 - \frac{1}{2}(s-1)\epsilon_2$, with $r,s=1,2,3,\ldots$, and $\epsilon_1,\epsilon_2$ are simply related to parameters $\beta,g_s$ (and encode the Omega-background in gauge theory interpretation). Furthermore, these differential equations have the structure of singular vectors of the relevant symmetry algebra, and can be expressed in terms of relevant representation of the symmetry generators.

Our results lead to several other interesting conclusions. First, we find interesting universal expressions for singular vectors, which do not seem to have been known before. For a fixed level, these expressions depend on the parameter $\alpha$, and setting this value to be equal to the value of the degenerate momentum at the same or any lower level, these expressions reduce to appropriate expressions for singular vectors corresponding to the given degenerate momentum. Second, we note that there are two interesting classical limits of the quantum curves that we consider: the 't Hooft limit of large size of matrices $N$ (or equivalently vanishing $g_s$), and the classical limit of infinite central charge from the viewpoint of Liouville conformal field theory, which is equivalent to the Nekrasov-Shatashvili limit and corresponds to vanishing or infinite value of $\beta$. In this sense the quantum curves that we consider can be interpreted as doubly-quantized objects, similarly (and especially after replacing parameters $g_s$ and $\beta$ by $\epsilon_1$ and $\epsilon_2$) to considerations in the context of the Langlands correspondence in \cite{Teschner:2010je}. Third, upon specialization of our analysis to multi-Penner or super-multi-Penner models, corresponding quantum curves and other objects reduce to expressions familiar from conformal field theory studies, such as (supersymmetric) BPZ equations, Ward identities, correlation functions and a representation of (super-)Virasoro operators acting thereon, etc.

We also note that results summarized in this note -- at least in the context of the Virasoro algebra and standard $\beta$-deformed matrix models -- can be rephrased in the language of M-theory, surface operators, as well as refined topological string theory (more precisely, the $\beta$-deformed hermitian matrix model can be regarded as a definition of refined topological string theory \cite{Dijkgraaf:2009pc,ACDKV}, whose worldsheet formulation has not been found to date). In the M-theory interpretation, one considers a background geometry of the form
\be
S^1\times \textrm{Taub-NUT}\times X,   \label{Mtheory}
\ee
where $X$ is a Calabi-Yau threefold mirror to the geometry $A(x,y)=uv$, with $A(x,y)$ given in (\ref{Axy}). The Taub-NUT space is parametrized by complex parameters $z_1$ and $z_2$, and twisted such that a rotation along the circle $S^1$ induces a rotation $z_1\mapsto e^{i\epsilon_1}z_1, z_2\mapsto e^{i\epsilon_2}z_2$. In addition, in this background one can introduce M5-branes that wrap $S^1$, a lagrangian subspace of $X$, and a complex line in the Taub-NUT space parametrized by $z_1$ or $z_2$. These M5-branes are referred to as $\epsilon_1$-branes and $\epsilon_2$-branes respectively. The quantum curves and wave-functions labeled by Virasoro degenerate momentum $\alpha=\alpha_{r,s}$ (with $r,s=1,2,3,\ldots$) represent a stack that consists of $(s-1)$ overlapping $\epsilon_1$-branes and $(r-1)$ overlapping $\epsilon_2$-branes. Reducing this system along $X$ leads to analogous interpretation with a stack of $(s-1)$ surface operators of one type and $(r-1)$ surface operators of another type in four or five-dimensional supersymmetric gauge theory, with non-trivial Omega-background parametrized by $\epsilon_1$ and $\epsilon_2$. On the other hand, reducing this system to the internal space and taking the mirror leads to the refined B-model topological string configuration on the background $A(x,y)=uv$, with two types of B-branes, whose wave-function is also labeled by $\alpha=\alpha_{r,s}$. Such systems with a single $\epsilon_1$-brane or a single $\epsilon_2$-brane are discussed in \cite{ACDKV}.

The plan of this note is as follows. In section \ref{sec-Virasoro} we present the singular vector structure of quantum curves in the bosonic case, with the underlying Virasoro algebra. In section \ref{sec-super} we discuss super-eigenvalue models and introduce super-quantum curves, which have the structure of singular vectors of the super-Virasoro algebra. Our presentation is based in particular on \cite{Manabe:2015kbj,Ciosmak:2016wpx}; however we mostly present main statements and results, and recommend the original papers to those readers who wish to get acquainted with proofs and technical computations. Section \ref{sec-summary} includes some comments and closing remarks.


\section{Higher level (Virasoro) quantum curves}     \label{sec-Virasoro}

In this section we determine higher level quantum curves corresponding to singular vectors of the Virasoro algebra, by considering $\beta$-deformation of the hermitian one-matrix model; we closely follow \cite{Manabe:2015kbj}, where more details and proofs are presented.


\subsection{$\alpha/\beta$-deformed matrix integrals}

To start with we introduce an expression that we refer to as $\alpha/\beta$-deformed matrix integral, which will play role of the wave-function
\be
\widehat{\psi}_{\alpha}(x) = \frac{e^{-\frac{2\alpha}{\epsilon_1 \epsilon_2}V(x)}}{(2\pi)^NN!}  \int_{{\IR}^N}   \Delta(z)^{2\beta}
\, \Big( \prod_{a=1}^N(x-z_a)^{-\frac{2\alpha}{\epsilon_{2}}} \Big)\,
e^{-\frac{\sqrt{\beta}}{\hbar}\sum_{a=1}^NV(z_a)} \prod_{a=1}^Ndz_a.    \label{Psi-alpha}
\ee
Ignoring the overall normalization and the term in the bracket under the integral (or simply for $\alpha=0$), this expression reduces to the eigenvalue representation of the partition function $Z\equiv\widehat{\psi}_{\alpha=0}(x)$ of the standard $\beta$-deformed matrix model. By $\Delta(z)=\prod_{1\le a<b\le N}(z_a-z_b)$ we denote the Vandermonde determinant, and various parameters that we use are related as follows
\begin{equation}
\epsilon_1=-\beta^{1/2}g_s,\qquad \epsilon_2=\beta^{-1/2}g_s,\qquad g_s=2\hbar,\qquad b^2 = -\beta  = \frac{\epsilon_1}{\epsilon_2}.
\label{matparam-intro}
\end{equation}
The exponential term under the integral is the eigenvalue representation of $e^{-\frac{\sqrt{\beta}}{\hbar}\textrm{Tr}V(M)}$, with the potential $V=V(x)$ parametrized by times $t_n$ and assumed to take form
\be
V(x) = \sum_{n=0}^{\infty}  t_n x^n.   \label{V}
\ee

For  $\beta=1,\frac{1}{2},2$ the expression (\ref{Psi-alpha}) is an eigenvalue representation of an integral over respectively hermitian, orthogonal, and symplectic matrices $M$. The term in the bracket under the integral is the eigenvalue representation of the determinant $\textrm{det}(x-M)$, raised to a power parametrized by $\alpha$. To the above model one can also associate the chiral boson field $\phi(x)$, with the background charge $Q = i\big(b+\frac{1}{b}\big) = \frac{\epsilon_1+\epsilon_2}{g_s}$ and central charge $c=1-6Q^2$
\begin{equation}
\phi(x)=-\frac{g_s N}{\epsilon_2}\log x + \frac{1}{g_s}\sum_{n=0}^{\infty}t_n x^n
-\frac{g_s}{2} \sum_{n=1}^{\infty} \frac{1}{nx^n}\partial_{t_n} 
=\frac{1}{g_s}V(x)-\frac{g_s}{\epsilon_2}\sum_{a=1}^N\log (x-z_a),   \label{phi-field}
\end{equation}
where we identified derivatives with respect to times of the partition function $Z$ (or expectation values of time-independent operators) with unnormalized expectation values (see (\ref{v_exp_d}) for the definition)
\begin{equation}
\Big\langle \sum_{a=1}^N z_a^n \, \cdots  \Big\rangle =
-\frac{\epsilon_2}{2}\partial_{t_n}  \Big\langle \cdots \Big\rangle.    \label{dt=z_a}
\end{equation}
The wave-function (\ref{Psi-alpha}) arises then as the expectation value of the exponent $e^{\frac{2\alpha}{g_s}\phi(x)}$, which in particular gives rise to the prefactor $e^{-\frac{2\alpha}{\epsilon_1 \epsilon_2}V(x)}$ (and keeping it simplifies some calculations).


\subsection{Higher level quantum curves as Virasoro singular vectors}

The main question that we ask is whether a putative wave-function (\ref{Psi-alpha}) satisfies a finite order differential equation in parameter $x$. In \cite{Manabe:2015kbj} we show that this is so only for discrete values of $\alpha$ of the form
\be
\alpha = \alpha_{r,s} = -\frac{r-1}{2}\epsilon_1 - \frac{s-1}{2}\epsilon_2,  \qquad \textrm{for}\ r,s=1,2,3,\ldots  \label{alpha-rs}
\ee
These values agree with degenerate momenta of the chiral boson in presence of the background charge, and up to a normalization by $ig_s$ can be written in the form $(1-r)b+(1-s)b^{-1}$. Furthermore, we find that for a particular value of $\alpha=\alpha_{r,s}$ the wave-function $\widehat{\psi}_{\alpha}(x)$ satisfies a differential equation of order $n=rs$ in $x$, which we write as
\be
\widehat{A}^{\alpha}_n \widehat{\psi}_{\alpha}(x) = 0.  \label{Aalpha-psialpha-intro}
\ee
We refer to operators $\widehat{A}^{\alpha}_n$ as (higher level) quantum curves. In general they take form
\be
\widehat{A}^{\alpha}_n = \sum_{p_1+p_2+\ldots+p_k=n}  \widehat{c}_{p_1,p_2,\ldots,p_k}(\alpha)\,   \widehat{L}_{-p_1} \widehat{L}_{-p_2}\cdots \widehat{L}_{-p_k},       \label{chat-Lhat-intro}
\ee
where operators $\widehat{L}_{-p}$ form a representation of the Virasoro algebra
\be
[L_m,L_n]=(m-n)L_{m+n}+\frac{c}{12}(m^3-m)\delta_{m+n,0}.
\ee
The crucial observation is that (\ref{chat-Lhat-intro}) has the same form as an operator $A_{r,s}$ that acting on a primary state $|\Delta_{r,s}\rangle$ creates a Virasoro singular vector corresponding to a given value $\alpha_{r,s}$. The constants $\widehat{c}_{p_1,p_2,\ldots,p_k}(\alpha)$, after specialization to $\alpha=\alpha_{r,s}$, have the same values as in known expressions for singular vectors. Indeed, in general, Virasoro singular vectors (at level $n=rs$) are labeled by two positive integers $r,s$ (that also label corresponding degenerate momenta (\ref{alpha-rs})) and take form $A_{r,s}|\Delta_{r,s}\rangle$, where $|\Delta_{r,s}\rangle$ is the primary state of weight $\Delta_{r,s}= \frac{1-r^2}{4}b^2 + \frac{1-s^2}{4}b^{-2} + \frac{1-rs}{2}$; the operators $A_{r,s}$ for the first few singular vectors take form
\begin{align}
A_{2,1} &= L_{-1}^2 + b^2 L_{-2}  \nonumber  \\
A_{3,1} &= L_{-1}^3  + 2b^2 L_{-1}L_{-2} + 2b^2 L_{-2}L_{-1} + 4b^4 L_{-3}
= L_{-1}^3  + 4b^2 L_{-2}L_{-1} + (2b^2 + 4b^4 )L_{-3}  \nonumber   \\
A_{4,1} & = L_{-1}^4 + 10b^2 L_{-1}^2 L_{-2} + 9b^4L_{-2}^2 + (24b^4-10b^2)L_{-1}L_{-3} + (36b^6 - 24b^4 + 6b^2)L_{-4} \nonumber \\
A_{2,2} &= L_{-1}^4 + \big(b^2-b^{-2}\big)^2 L_{-2}^2 + \frac{3}{2}\big(b+b^{-1}  \big)^2 L_{-1}^2L_{-2} +
\frac{3}{2}\big(b-b^{-1}  \big)^2 L_{-2}  L_{-1}^2 + \nonumber \\
&\qquad - \big( b^2 + b^{-2}\big) L_{-1}L_{-2}L_{-2}      \label{null-level234}
\end{align}
and operators $A_{s,r}$ are related to $A_{r,s}$ by replacing $b$ by $b^{-1}$.

It is noteworthy that expressions of the form (\ref{chat-Lhat-intro}) that we find (and which do not seem to have been known before) have general, universal dependence on $\alpha$, and our work provides an algorithm to determine coefficients $\widehat{c}_{p_1,p_2,\ldots,p_k}(\alpha)$ that encode at once all singular vectors up to a given level; such coefficients determined explicitly up to level 5 are summarized below. This result is interesting, especially that, even though it is known that Virasoro singular vectors can be written in terms of Jack polynomials \cite{mimachi1995}, determining their explicit form is still an important problem in conformal field theory (which is solved only in some particular cases, such as $r=1$ or $s=1$ \cite{BenoitSaintAubin}).

Furthermore, the operators $\widehat{L}_{-p}$ in (\ref{chat-Lhat-intro}) with $p\geq 0$ form the following interesting representation of the Virasoro algebra on a space of functions in $x$ and times $t_k$
\be
\begin{split}
\widehat{L}_0\,  &=\Delta_{\alpha}\equiv \frac{\alpha}{g_s}\Big(\frac{\alpha}{g_s} - Q \Big) ,  \qquad  \qquad   \widehat{L}_{-1} =\partial_x ,  \label{Lm_fer_op-intro}  \\
\, \widehat{L}_{-n} = - \frac{1}{\epsilon_1\epsilon_2 (n-2)!} \Big( &  \partial_x^{n-2} \big(  V'(x)^2  \big)  + (\epsilon_1+\epsilon_2) \partial_x^n V(x)  + \partial_x ^{n-2}\hat{f}(x) \Big),\  \textrm{for}\ n\geq 2, \,
\end{split}
\ee
where $\partial_x^n\widehat{f}(x) \equiv [\partial_x, \partial_x^{n-1}\widehat{f}(x)]$ and
\begin{equation}
\hat{f}(x)=-\epsilon_1\epsilon_2\sum_{m=0}^{\infty}x^m\partial_{(m)},\ \ \ \
\partial_{(m)}=\sum_{k=m+2}^{\infty}kt_k\frac{\partial}{\partial t_{k-m-2}}.
\label{sp_hat_f_op-intro}
\end{equation}
The operator $\hat{f}(x)$ acting on the partition function $Z$ has the same effect as computation of the expectation value of the expression
\be
f(x)=2\epsilon_1\sum_{a=1}^N\frac{V'(x)-V'(z_a)}{x-z_a}.  \label{fx}
\ee
The representation (\ref{Lm_fer_op-intro}) can be found by interpreting (\ref{Psi-alpha}) in conformal field theory language, as an insertion of an operator $\prod_a (x-z_a)^{-\frac{2\alpha}{\epsilon_{2}}}$ at a position $x$, and determining modes of the corresponding energy-momentum tensor. As $\widehat{L}_{-n}$ involves derivatives with respect to times, encoded in the term $\partial_x ^{n-2}\hat{f}(x)$, it follows that operators (\ref{chat-Lhat-intro}) are time-dependent quantum curves, so in general they impose partial differential equations in $x$ and times $t_k$. However, for certain choices of the potential $V(x)$ -- e.g. for Gaussian or multi-Penner potentials -- quantum curves become time-independent and impose ordinary differential equations in $x$ for $\widehat{\psi}_{\alpha}(x)$.

Apart from (\ref{Psi-alpha}), in certain situations it is useful to consider the wave-function normalized by the partition function $Z$
\be
\Psi_{\alpha} = \frac{\widehat{\psi}_{\alpha}(x)}{Z}.   \label{PsiPsi-alpha}
\ee
Taking advantage of the relation $Z^{-1}\widehat{L}_{-n}\widehat{\psi}_{\alpha}(x) =Z^{-1}\widehat{L}_{-n}Z\widehat{\Psi}_{\alpha}(x) \equiv \widehat{\mathcal L}_{-n}\widehat{\Psi}_{\alpha}(x)$, when acting on $\Psi_{\alpha} $, Virasoro generators (\ref{Lm_fer_op-intro}) are transformed into
\begin{align}
\widehat{\mathcal L}_{-n}&=Z^{-1}\widehat{L}_{-n}Z =   \label{calLn} \\
&=
-\frac{1}{\epsilon_1\epsilon_2(n-2)!}\Big(\partial_x^{n-2}\big(V'(x)^2\big)+(\epsilon_1+\epsilon_2)\partial^n_x V(x)
+\partial_x^{n-2}\widehat{f}(x) + \big[\partial_x^{n-2}\widehat{f}(x), \log Z \big]\Big).    \nonumber
\end{align}



In the rest of this section we present explicit expressions for higher level quantum curves, up to level 5. Imposing the condition that $\widehat{\psi}_{\alpha}(x)$ satisfies a second order differential equation in $x$, we find that quantum curves at level 2 take form
\be
\widehat{A}^{\alpha}_2 \widehat{\psi}_{\alpha}(x) \equiv \Big(\widehat{L}_{-1}^2+\frac{4\alpha^2}{\epsilon_1\epsilon_2}\widehat{L}_{-2}\Big) \widehat{\psi}_{\alpha}(x)=0,\ \ \
\mbox{for}\ \alpha=-\frac{\epsilon_1}{2}, -\frac{\epsilon_2}{2}.       \label{BPZ-level2-intro}
\ee
For the special values $\alpha=\alpha_{2,1}=-\frac{\epsilon_1}{2}$ or $\alpha=\alpha_{1,2}=-\frac{\epsilon_2}{2}$, the above differential operator takes form $\widehat{A}^{\alpha}_2=\widehat{L}_{-1}^2+b^{\pm 2} \widehat{L}_{-2}$ respectively, with $b^2$ defined in (\ref{matparam-intro}). This indeed agrees with well known expressions for operators that create Virasoro singular vectors at level 2, see (\ref{null-level234}), and analogous calculation was first presented in \cite{ACDKV}. We stress that the form (\ref{BPZ-level2-intro}) is universal, and reduces to either of the two singular vectors at level 2 upon appropriate choice of $\alpha$. In terms of the representation (\ref{Lm_fer_op-intro}), quantum curves in (\ref{BPZ-level2-intro}) have an explicit form
\begin{equation}
\widehat{A}^{\alpha}_2 = \partial_x^2-\frac{4\alpha^2}{\epsilon_1^2\epsilon_2^2}V'(x)^2-\frac{4\alpha^2}{\epsilon_1^2\epsilon_2^2}(\epsilon_1+\epsilon_2)V''(x)
-\frac{4\alpha^2}{\epsilon_1^2\epsilon_2^2}\widehat{f}(x).\ \ \
\label{BPZ-level-2bis-intro}
\end{equation}

Quantum curves at higher levels can be found analogously, by demanding that $\widehat{\psi}_{\alpha}(x)$ satisfies a differential equation in $x$ of some particular order. At level 3 we find the result
\be
\widehat{A}_3^{\alpha}  = \widehat{L}_{-1}\widehat{A}_2^{\alpha}
+\frac{2\alpha^2}{\epsilon_1^2\epsilon_2^2}(2\alpha+\epsilon_1)(2\alpha+\epsilon_2)\widehat{L}_{-3}.    \label{BPZ-level-3-intro} \\ 
\ee
For $\alpha=\alpha_{3,1}=-\epsilon_1$ or $\alpha=\alpha_{1,3}=-\epsilon_2$ this operator reduces to known expressions (\ref{null-level234}) for singular vectors at level 3; in our context they can be written down more explicitly, using the representation (\ref{Lm_fer_op-intro}). Furthermore, substituting values of $\alpha$ corresponding to level 2, i.e. $\alpha_{2,1}=-\frac{\epsilon_1}{2}$ or $\alpha_{1,2}=-\frac{\epsilon_2}{2}$, the second term in (\ref{BPZ-level-3-intro}) vanishes and $\widehat{A}_3^{\alpha}$ reduces to quantum curves $\widehat{A}_2^{\alpha}$ at level 2. In this sense the expression (\ref{BPZ-level-3-intro}) encodes all singular vectors up to and including level 3. Furthermore, quantum curve at level 4 takes form
\be
\begin{split}
\ \widehat{A}_4^{\alpha}&=
\widehat{L}_{-1}\widehat{A}_3^{\alpha}
+\frac{4\alpha(\alpha+\epsilon_1)(\alpha+\epsilon_2)}{\epsilon_1\epsilon_2(5\alpha+3\epsilon_1+3\epsilon_2)}\widehat{L}_{-2}\widehat{A}_2^{\alpha}\, + \label{BPZ-level-4c}  \\  
&\ \
-\frac{2\alpha(2\alpha+\epsilon_1)(2\alpha+\epsilon_2)(\alpha+\epsilon_1)(\alpha+\epsilon_2)}
{\epsilon_1^3\epsilon_2^3(5\alpha+3\epsilon_1+3\epsilon_2)}
\Big(\epsilon_1\epsilon_2\widehat{L}_{-1}\widehat{L}_{-3}
-2(2\alpha+\epsilon_1)(2\alpha+\epsilon_2)\widehat{L}_{-4}\Big).
\end{split}
\ee
This expression reduces to known expressions (\ref{null-level234}) for singular vectors at level 4 upon substitution of the relevant values of degenerate momenta $\alpha_{4,1}=-\frac{3\epsilon_1}{2}$, $\alpha_{1,4}=-\frac{3\epsilon_2}{2}$, $\alpha_{2,2}=-\frac{\epsilon_1+\epsilon_2}{2}$. For degenerate momenta corresponding to lower levels, (\ref{BPZ-level-4c}) factorizes and reduces to expressions for singular vectors at those lower levels.

Taking advantage of the following notation
\begin{align}
\begin{split}
&
\gamma_1=\frac{\alpha^2}{\epsilon_1\epsilon_2},\qquad
\gamma_2=\frac{(2\alpha+\epsilon_1)(2\alpha+\epsilon_2)}{\epsilon_1\epsilon_2},\qquad
\gamma_3=\frac{(\alpha+\epsilon_1)(\alpha+\epsilon_2)}{\epsilon_1\epsilon_2},\\
&
\gamma_4=\frac{(2\alpha+3\epsilon_1)(2\alpha+3\epsilon_2)(2\alpha+\epsilon_1+\epsilon_2)}{\epsilon_1\epsilon_2\alpha},
\\
&
\delta_1=\frac{\alpha}{5\alpha+3\epsilon_1+3\epsilon_2},\qquad
\delta_2=\frac{\alpha}{7\alpha+6\epsilon_1+6\epsilon_2},
\end{split}
\end{align}
quantum curves at level 5, and also all lower levels, can be written compactly in the form
\begin{align}
\begin{split}
\widehat{A}_2^{\alpha}&=\widehat{L}_{-1}^2+4\gamma_1\widehat{L}_{-2},
\\
\widehat{A}_3^{\alpha}&=\widehat{L}_{-1}\widehat{A}_2^{\alpha}
+2\gamma_1\gamma_2\widehat{L}_{-3},
\\
\widehat{A}_4^{\alpha}&=
\widehat{L}_{-1}\widehat{A}_3^{\alpha}
+4\delta_1\gamma_3\widehat{L}_{-2}\widehat{A}_2^{\alpha}
-2\delta_1\gamma_2\gamma_3
\big(\widehat{L}_{-1}\widehat{L}_{-3}
-2\gamma_2\widehat{L}_{-4}\big),  \label{BPZ-levels-234}
\\
\widehat{A}_5^{\alpha} &= \widehat{L}_{-1}\widehat{A}_4^{\alpha}
+2\delta_1\delta_2\gamma_4\big(2\widehat{L}_{-2}\widehat{A}_3^{\alpha} + \gamma_3\widehat{L}_{-3}\widehat{A}_2^{\alpha}\big)
-4\delta_2\gamma_2\gamma_3\gamma_4
\big(\delta_1\widehat{L}_{-1}\widehat{L}_{-4} - (\gamma_1+3\delta_1)\widehat{L}_{-5}\big).\
\end{split}
\end{align}
In particular, setting $\alpha$ to be equal to $\alpha_{5,1}=-2\epsilon_1$ or $\alpha_{1,5}=-2\epsilon_2$, the operator $\widehat{A}_5^{\alpha}$ reduces to known expressions for singular vectors at level 5, and substituting values of $\alpha$ corresponding to lower levels this operator reduces to the form of singular vectors at those lower levels.


With some effort one can find analogous universal, $\alpha$-dependent expressions for quantum curves and Virasoro singular vectors at higher levels. Simplifying the algorithm presented in \cite{Manabe:2015kbj} that enables to find such higher level quantum curves, and determining their form explicitly at arbitrary level, are important tasks for future work.


\subsection{Double quantum structure}    \label{ssec-double}

The wave-functions and quantum curves that we consider, apart from a discrete value of $\alpha$, depend on two continuous parameters $g_s$ and $\beta$, or equivalently $\epsilon_1$ and $\epsilon_2$ defined in (\ref{matparam-intro}). It turns out that one can consider two corresponding classical limits, namely 't Hooft large $N$ (or vanishing $g_s$) limit, and the classical CFT limit of an infinite central charge. It follows that quantum curves that we consider can be regarded as doubly quantized objects, analogously to the discussion in the context of the Langlands correspondence in \cite{Teschner:2010je}. Let us briefly discuss these two quantization procedures. Note that to obtain a well defined, non-divergent classical limit, one has to subtract the partition function $Z$ from the wave-function; therefore in this section we consider quantum curves and wave-functions $\Psi_{\alpha}(x)$ in the normalization (\ref{PsiPsi-alpha}).

To start with, we consider the classical 't Hooft limit for quantum curves at level 2, for $\beta=1$. In this case (\ref{BPZ-level-2bis-intro}), for either value $\alpha=-\epsilon_1/2$ or $-\epsilon_2/2$, reduces to
\be
\Big( g_s^2 \partial_x^2 - V'(x)^2 -\widehat{f}(x)\Big)\widehat{\psi}_{\alpha}(x) = 0.
\ee
Dividing this equation by the partition function $Z$ we obtain a differential equation for $\Psi_{\alpha}$, and using the factorization of expectation values, in the large $N$ limit we find the equation
\be
y^2 - V'(x)^2 - f_{cl}(x) = 0,   \label{Ahat-class-limit-level2}
\ee
where $y$ is identified with the classical limit of $g_s\partial_x$, while
\be
f_{cl}(x) = \lim_{N\to\infty}  \frac{\langle f(x) \rangle}{Z}.   \label{f-cl}
\ee
The result (\ref{Ahat-class-limit-level2}) is precisely the spectral curve of the hermitian one-matrix model (as defined by (\ref{Psi-alpha}) with $\alpha=0$ and $\beta=1$). It follows that the quantum curve (\ref{BPZ-level-2bis-intro}) can be regarded as the quantization of the spectral curve.

Similarly one can analyze the classical 't Hooft limit for quantum curves at higher level. As shown in \cite{Manabe:2015kbj}, in this limit $\widehat{L}_{-p}$ with $p\geq 3$ can be set to zero, and $\widehat{L}_{-1}$ and $\widehat{L}_{-2}$ become commuting, which is apparently the same limit as analyzed in \cite{Feigin:1988se,Kent:1991wm}. In this limit resulting classical curve equations factorize. In particular, the curves corresponding to $\alpha_{r,1}$, reduce to
\be
\begin{split}
0 &= \prod_{k=1}^{r/2}  \Big( y^2 - \frac{(2k-1)^2}{(r-1)^2} \big(V'(x)^2 + f_{cl}(x) \big)  \Big)  ,\ \quad \qquad \textrm{for $r$ even}     \\
0 &= y \prod_{k=1}^{(r-1)/2}  \Big( y^2 - \frac{4k^2}{(r-1)^2} \big(V'(x)^2 + f_{cl}(x) \big)  \Big),\ \, \qquad \textrm{for $r$ odd}    \label{class-curves-higher-level}
\end{split}
\ee
where $y$ is identified with the limit of $\frac{\epsilon_2}{r-1}\partial_x$. Each factor in the above expressions represents the spectral curve (\ref{Ahat-class-limit-level2}), with $(V'(x)^2 + f_{cl}(x))$ term (or simply $y$) rescaled by a simple factor.  In other words, a product of several classical curves is non-trivially resolved in the process of quantization, in a way which is consistent with the structure of singular vectors, and quantum curves at higher levels can be regarded as non-trivial resolutions of multiple copies of the underlying spectral curve.

The second classical limit that we consider is that of infinite central charge, analogous to the classical limit in Liouville theory. This is also equivalent to the Nekrasov-Shatashvili limit, where either $\epsilon_1$ or $\epsilon_2$ is set to zero. For definiteness we choose the case $\epsilon_1\to 0$, which is equivalent to vanishing $b$, as follows from the relation $b^2=\frac{\epsilon_1}{\epsilon_2}$. In terms of $b$ and $g_s$ parameters, in order to keep $\epsilon_2$ constant we need to consider a double scaling limit, with both of these parameters vanishing with a constant ratio. In the limit $\epsilon_1\to 0$ it is natural to consider quantum curves and wave-functions labeled by
\begin{equation}
\alpha=\alpha_{r+1,1}=-\frac{r}{2}\epsilon_1.
\label{p_ep_2_br}
\end{equation}
As mentioned above, we consider wave-functions normalized by the partition function (\ref{PsiPsi-alpha}), which in the $\epsilon_1\to 0$ limit we denote by $\Psi^{\textrm{NS}}_{\alpha}(x)$, and which factorize in this limit as follows
\be
\Psi^{\textrm{NS}}_{-\frac{r}{2}\epsilon_1}(x)
\equiv \lim_{\epsilon_1\to 0}\Psi_{-\frac{r}{2}\epsilon_1}(x)
=\left(\Psi^{\textrm{NS}}_{-\frac{1}{2}\epsilon_1}(x)\right)^r.
\ee
These wave-functions are annihilated by operators that arise in the limit of (\ref{chat-Lhat-intro})
\be
\widehat{\mathcal{A}}_{r+1}^{\textrm{NS}}  \Psi^{\textrm{NS}}_{-\frac{r}{2}\epsilon_1}(x) = 0, \qquad\quad
\widehat{\mathcal{A}}_{r+1}^{\textrm{NS}} = \lim_{\epsilon_1\to 0} \epsilon_2^{r+1}Z^{-1}\widehat{A}_{r+1}^{-\frac{r}{2}\epsilon_1}Z.
\end{equation}
The operators $\widehat{\mathcal{A}}_{r+1}^{\textrm{NS}}$ have the structure of singular vectors, written in terms of Virasoro generators that arise in the appropriate limit of  (\ref{calLn})
\begin{equation}
\widehat{\mathcal L}_{-n}^{\textrm{NS}}=\lim_{\epsilon_1\to 0} \epsilon_1\epsilon_2 \widehat{\mathcal L}_{-n}
= -\frac{1}{(n-2)!}\Big(\partial_x^{n-2}\big(V'(x)^2\big)+\epsilon_2
\partial_x^n V(x)+F^{(0)}_{n-2}(x,\epsilon_2)\Big),
\end{equation}
where  $F^{(0)}_{0}(x,0)=f_{cl}(x)$ given in (\ref{f-cl}), and more generally we defined
\begin{equation}
F^{(0)}(\epsilon_2)=-\lim_{\epsilon_1\to 0}\epsilon_1\epsilon_2\log Z, \qquad\quad
F^{(0)}_{k}(x,\epsilon_2)=\sum_{n=k}^{\infty}\frac{n!}{(n-k)!}x^{n-k}\partial_{(n)}F^{(0)}(\epsilon_2).
\end{equation}
For example, in the $\epsilon_1\to 0$ limit the quantum curve equation (\ref{BPZ-level2-intro}) at level 2 reduces to
\begin{equation}
\widehat{\mathcal{A}}_2^{\textrm{NS}}\Psi^{\textrm{NS}}_{-\frac{1}{2}\epsilon_1}(x)
= \Big(\epsilon_2^2\partial_x^2+\widehat{\mathcal L}^{\textrm{NS}}_{-2}\Big)\Psi^{\textrm{NS}}_{-\frac{1}{2}\epsilon_1}(x)=0.
\end{equation}

Now it is useful to define inductively  differential operators $\widehat{a}_{q}^{r+1}$ for $q=0,1,\ldots,r+1$
\begin{equation}
\widehat{a}_{0}^{r+1}=1,\ \ \
\widehat{a}_{1}^{r+1}=\epsilon_2\partial_x,\ \ \
\widehat{a}_{q+1}^{r+1}=\epsilon_2\partial_x\widehat{a}_{q}^{r+1}
+q(r-q+1)\widehat{\mathcal L}^{\textrm{NS}}_{-2}\widehat{a}_{q-1}^{r+1}.
\label{NS_B_op}
\end{equation}
By induction one can show that
\begin{equation}
\widehat{a}_{q+1}^{r+1}\Psi^{\textrm{NS}}_{-\frac{r}{2}\epsilon_1}(x)=
r(r-1)(r-2)\cdots (r-q)\left(\Psi^{\textrm{NS}}_{-\frac{1}{2}\epsilon_1}(x)\right)^{r-q-1}
\left(\epsilon_2\partial_x \Psi^{\textrm{NS}}_{-\frac{1}{2}\epsilon_1}(x)\right)^{q+1}.
\end{equation}
Therefore the wave-function $\Psi^{\textrm{NS}}_{-\frac{r}{2}\epsilon_1}(x)$ satisfies
an ordinary differential equation of order $(r+1)$, and the corresponding quantum curve can be identified as
$\widehat{\mathcal{A}}_{r+1}^{\textrm{NS}} = \widehat{a}_{r+1}^{r+1}$.
In particular
\be
\begin{split}
\widehat{\mathcal{A}}_{2}^{\textrm{NS}}&= \epsilon_2^2\partial_x^2+\widehat{\mathcal L}^{\textrm{NS}}_{-2},  \\
\widehat{\mathcal{A}}_{3}^{\textrm{NS}}&=
\epsilon_2^3\partial_x^3+4\epsilon_2\widehat{\mathcal L}^{\textrm{NS}}_{-2}\partial_x+2\epsilon_2\widehat{\mathcal L}^{\textrm{NS}}_{-3},
\\
\widehat{\mathcal{A}}_{4}^{\textrm{NS}}&=
\epsilon_2^4\partial_x^4
+10\epsilon_2^2\widehat{\mathcal L}^{\textrm{NS}}_{-2}\partial_x^2
+10\epsilon_2^2\widehat{\mathcal L}^{\textrm{NS}}_{-3}\partial_x
+9\big(\widehat{\mathcal L}^{\textrm{NS}}_{-2}\big)^2
+6\epsilon_2^2\widehat{\mathcal L}^{\textrm{NS}}_{-4},
\\
\widehat{\mathcal{A}}_{5}^{\textrm{NS}}&=
\epsilon_2^5\partial_x^5
+20\epsilon_2^3\widehat{\mathcal L}^{\textrm{NS}}_{-2}\partial_x^3
+30\epsilon_2^3\widehat{\mathcal L}^{\textrm{NS}}_{-3}\partial_x^2
+64\epsilon_2\big(\widehat{\mathcal L}^{\textrm{NS}}_{-2}\big)^2\partial_x
+36\epsilon_2^3\widehat{\mathcal L}^{\textrm{NS}}_{-4}\partial_x +  \\
&\ \ \
+64\epsilon_2\widehat{\mathcal L}^{\textrm{NS}}_{-2}\widehat{\mathcal L}^{\textrm{NS}}_{-3}
+24\epsilon_2^3\widehat{\mathcal L}^{\textrm{NS}}_{-5}.
\end{split}
\ee
Using the Virasoro algebra
$\partial_x^n \widehat{\mathcal L}^{\textrm{NS}}_{-2}  = n! \widehat{\mathcal L}^{\textrm{NS}}_{-n-2}$, each $\widehat{\mathcal{A}}_{r+1}^{\textrm{NS}}$ can also be expressed in terms of $\partial_x$ and (derivatives of) $\widehat{\mathcal L}^{\textrm{NS}}_{-2}$ only. If we further identify the energy-momentum tensor in classical Liouville theory as $T^{(c)}\equiv \widehat{\mathcal L}^{\textrm{NS}}_{-2}$, then the operators $\widehat{\mathcal{A}}_{r+1}^{\textrm{NS}}$ take the same form as operators imposing differential equations for the fields $e^{-r\varphi/2}$ in the classical Liouville theory \cite{Zamolodchikov:2003yb}. Furthermore, taking the limit $\epsilon_2\to 0$, the results we just obtained reduce to those in the 't Hooft limit discussed first.


\subsection{Examples: Gaussian and multi-Penner models}    \label{ssec-examples}

In this section we illustrate our general considerations in two examples, of Gaussian and multi-Penner models. One interesting feature is that in these cases quantum curves can be reduced to ordinary (not partial) differential equations. Second, multi-Penner model is particularly interesting, as in this case our formalism reduces essentially to various results familiar from conformal field theory.

We discuss first the Gaussian model, i.e. the model defined by (\ref{Psi-alpha}) with the quadratic potential, which is a specialization of (\ref{V}) to only one non-zero time $t_2=\frac{1}{2}$
\be
V(x) = \frac{1}{2} x^2.   \label{V-gaussian}
\ee
In this case (\ref{sp_hat_f_op-intro}) simplifies and its derivatives vanish
\be
\hat{f}(x)  = -\epsilon_1 \epsilon_2 \partial_{t_0}  , \qquad \qquad
 \partial_x^k \hat{f}(x)=0 \qquad\textrm{for} \ k\geq 1,
\ee
while operators (\ref{Lm_fer_op-intro}) (when acting on $\widehat{\psi}_{\alpha}(x)$) take form
\be
\widehat{L}_{-1} =\partial_x ,\qquad
\widehat{L}_{-2} = -\frac{x^2 +\epsilon_1+\epsilon_2 -4\mu + 2\alpha}{\epsilon_1\epsilon_2}, \qquad
\widehat{L}_{-3} = -\frac{2x}{\epsilon_1\epsilon_2}, \qquad
\widehat{L}_{-4} = -\frac{1}{\epsilon_1\epsilon_2},
\label{Lm_fer_op-gaussian}
\ee
where $\mu=\beta^{1/2}\hbar N$ is the 't Hooft parameter, and $\widehat{L}_{-n}=0$ for $n\geq 5$. In this way we obtain an interesting realization of a subalgebra of the Virasoro algebra -- truncated at $\widehat{L}_{-5}$ -- in terms of differential operators in one variable $x$. For polynomial potentials of higher degree one can  analogously obtain realization of larger Virasoro subalgebras, in terms of differential operators in several variables. The operators (\ref{Lm_fer_op-gaussian}) are building blocks of higher level quantum curves. For example, at level 2, for $ \alpha=-\frac{\epsilon_1}{2}, -\frac{\epsilon_2}{2} $ the equation (\ref{BPZ-level2-intro}) takes form
\be
\big(\widehat{L}_{-1}^2 + b^{\pm 2} \widehat{L}_{-2} \big) \widehat{\psi}_{\alpha}(x) =
\Big( \partial_x^2-\frac{4\alpha^2}{\epsilon_1^2\epsilon_2^2}
(x^2+\epsilon_1+\epsilon_2-4\mu+2\alpha)
 \Big)  \widehat{\psi}_{\alpha}(x) = 0 ,  \label{BPZ2-gaussian-pre}
\ee
where $b^2=\frac{\epsilon_1}{\epsilon_2}=-\beta$, and a choice of a sign in the exponent of $b^{\pm 2}$ corresponds respectively to the choice of $\alpha=-\frac{\epsilon_1}{2}$ or $\alpha= -\frac{\epsilon_2}{2}$. For $\alpha=-\frac{\epsilon_1}{2}$ or $\alpha= -\frac{\epsilon_2}{2}$, and in the unrefined limit $\beta=1$, from (\ref{BPZ2-gaussian-pre}) we obtain unrefined quantum curves
\be
\big(g_s^2\partial_x^2 -x^2 + 4\mu  \mp g_s \big) \widehat{\psi}_{\alpha}(x) =0, \label{BPZ2-gaussian}
\ee
and then (a unique) algebraic curve in the classical limit
\be
y^2 -x^2 + 4\mu = 0,       \label{gaussian-class-Psi}
\ee
with $y$ representing the classical limit of $g_s \partial_x$. At level 3, the unrefined limit of (\ref{BPZ-level-3-intro}) yields
\be
\Big( g_s^3\partial_x^3 - 4(x^2 -4\mu \pm2g_s)g_s\partial_x + 4g_s x \Big) \widehat{\psi}_{\alpha}(x) =0,
\label{BPZ3-gaussian}
\ee
where $\pm$ corresponds respectively to the choice of $\alpha=-\epsilon_1, -\epsilon_2$, and in the classical limit we find an algebraic curve which factorizes as
\be
2y\big((2y)^2-4x^2+16\mu\big) = 0,  \label{gaussian-class-Psi-level3}
\ee
in agreement with (\ref{class-curves-higher-level}) and the earlier discussion.

The second interesting example is the multi-Penner model, defined by the potential
\begin{equation}
V(x)=\sum_{i=1}^{M}\alpha_i\log (x-x_i).
\label{m_penner_potential}
\end{equation}
As is well known, this matrix model computes, in minimal models or in Liouville theory, correlation functions of $(M+1)$ primary fields with momenta $\alpha_i$ and $\alpha_{\infty}$, inserted respectively at positions $x_i$ and at infinity on ${\mathbb{P}}^1$ \cite{Dotsenko:1984nm,Mironov:2010zs,Mironov:2010su,Dijkgraaf:2009pc}. In what follows we show that various other objects that we introduced earlier, e.g. the representation of Virasoro operators $\widehat{L}_{n}$, higher level quantum curves, etc., reduce to familiar objects in minimal models or in Liouville theory for the potential (\ref{m_penner_potential}), with the dependence on an infinite set of times $t_n$ replaced by the dependence on parameters $x_i$, i.e. positions of operator insertions. We also recall that the multi-Penner model (\ref{m_penner_potential}) describes a four-dimensional ${\cal N}=2$ $SU(2)^{M-2}$ superconformal linear quiver gauge theory \cite{Dijkgraaf:2009pc}; for example for $M=2$ this model describes a theory of four free hypermultiplets. Via this connection, our results found from the perspective of matrix models can be interpreted in the language of supersymmetric gauge theories and related topics.

Note that for the potential (\ref{m_penner_potential}), the exponential term in the matrix model integrand (\ref{Psi-alpha}) takes form
\be
e^{-\frac{\sqrt{\beta}}{\hbar}\sum_{a=1}^NV(z_a)} = \prod_{i=1}^M\prod_{a=1}^N (z_a-x_i)^{-\frac{2\alpha_i}{\epsilon_2}}.
\ee
Each factor in this expression, corresponding to fixed $i$, has the same form as the determinant-like insertion that defines the wave-function $\widehat{\psi}_{\alpha}(x)$. Therefore for the potential (\ref{m_penner_potential}), the wave-function can be interpreted as a correlation function of $(M+2)$ primary fields in the presence of an additional field at infinity, or $(M+1)$ fields once the momentum conservation condition is imposed. This suggests that quantum curve equations in this case should reduce to differential equations for correlation functions of a number of primary fields, which include a distinguished field with a degenerate momentum $\alpha_{r,s}$ inserted at position $x$. These are nothing but the BPZ equations, and in what follows we show that they indeed arise from our general formalism. In particular, at level 2 we obtain in this way familiar in minimal models or Liouville theory hypergeometric differential  equations for four-point correlation functions.

For the potential (\ref{m_penner_potential}) we see that the operator $\widehat{f}(x)$ in (\ref{sp_hat_f_op-intro}) takes form
\begin{equation}
\widehat{f}(x)=-\epsilon_1\epsilon_2\sum_{i=1}^{M}\frac{1}{x-x_i}\partial_{x_i}.  \label{fhat-multiPenner}
\end{equation}
In what follows it is useful to introduce another normalization and consider the wave-function
\begin{equation}
\widetilde{\psi}_{\alpha}(x) = \widehat{\psi}_{\alpha}(x) \cdot \prod_{i\neq j}(x_i-x_j)^{-\frac{\alpha_i\alpha_j}{\epsilon_1\epsilon_2}}.
\label{m_penner_normalize}
\end{equation}
Then, rewriting (\ref{fhat-multiPenner}) accordingly, the representation of Virasoro algebra (\ref{Lm_fer_op-intro}), when acting on $\widetilde{\psi}_{\alpha}(x)$, takes form
$\widetilde{L}_0=\Delta_{\alpha}, \widetilde{L}_{-1}=\partial_x$, and
\be
\widetilde{L}_{-n}=
\sum_{i=1}^M\Big(\frac{(n-1)\Delta_{\alpha_i}}{(x_i-x)^n}
-\frac{1}{(x_i-x)^{n-1}}\partial_{x_i}\Big),
\quad \textrm{for } n \ge 2.
\label{vir_m_penner}
\ee
These generators coincide with well-known expressions for Virasoro generators acting on correlation functions in conformal field theory. These generators can be further simplified, taking advantage of $SL(2,\mathbb{C})$ invariance, which in the matrix model formulation gives rise to three additional equations that constrain the wave-function $\widetilde{\psi}_{\alpha}(x)$. These equations imply that three among $M$ partial derivatives $\partial_{x_i}$ in Virasoro generators $\widetilde{L}_{-n}$ can be expressed in terms $\partial_x$. In particular, for $M=2$, with the momentum conservation condition imposed, we find
\begin{equation}
\widetilde{L}_{-2}\widetilde{\psi}_{\alpha}(x)=
\bigg[-\sum_{i=1}^2 \frac{1}{x-x_i}\partial_x
+\sum_{i=1}^2 \frac{\Delta_{\alpha_i}}{(x-x_i)^2}
-\frac{\Delta_{\alpha_{1}}+\Delta_{\alpha_{2}}
+\Delta_{\alpha}}{(x-x_{1})(x-x_{2})}\bigg]\widetilde{\psi}_{\alpha}(x),   \label{L-2-mPenner-M2}
\end{equation}
while for $M=3$ we find
\begin{equation}
\widetilde{L}_{-2}\widetilde{\psi}_{\alpha}(x)=
\bigg[-\sum_{i=1}^3 \frac{1}{x-x_i}\partial_x
+\sum_{i=1}^3 \frac{\Delta_{\alpha_i}}{(x-x_i)^2}
+\sum_{I=1}^3 \frac{\Delta_{\alpha_I}-\Delta_{\alpha_{I+1}}-\Delta_{\alpha_{I+2}}
-\Delta_{\alpha}}{(x-x_{I+1})(x-x_{I+2})}\bigg]\widetilde{\psi}_{\alpha}(x).   \label{L-2-mPenner-M3}
\end{equation}
For both (\ref{L-2-mPenner-M2}) and (\ref{L-2-mPenner-M3}), other generators $\widetilde{L}_{-n}$ with $n>2$ can be similarly obtained from (\ref{vir_m_penner}).

We can now use the representation of Virasoro operators (\ref{vir_m_penner}) to construct quantum curves that annihilate the wave-function $\widetilde{\psi}_{\alpha}(x)$. Quantum curve equations in this case have the same form as the original BPZ equations in conformal field theory. We stress that these equations make sense only for degenerate values $\alpha=\alpha_{r,s}$ given in (\ref{alpha-rs}). Including an additional determinant-like insertion that defines the wave-function $\widetilde{\psi}_{\alpha}(x)$ itself, this wave-function is identified with a correlation function of $(M+1)$ fields with momenta $\alpha=\alpha_{r,s}$ and $\alpha_i$, $i=1,\ldots,M$, that are parameters of the potential (\ref{m_penner_potential}).

For $M=3$ quantum curves can be built from Virasoro operators (\ref{L-2-mPenner-M3}) and other relevant $\widetilde{L}_{-n}$. These quantum curves are written in terms of $\partial_x$ and no other derivatives, thus we obtain time-independent equations for the wave-function, which can be interpreted as the four-point function that includes a field with a degenerate momentum. At level 2 we find a hypergeometric differential equation, coinciding with the original BPZ equations at level 2 \cite{Belavin:1984vu}
\be
\big( \partial_x^2 + b^{\pm 2} \widetilde{L}_{-2} \big) \widetilde{\psi}_{\alpha}(x) = 0,
\ee
respectively for the choice of momentum $\alpha=\alpha_{2,1}$ and $\alpha=\alpha_{1,2}$, and
with $\widetilde{L}_{-2}$ given in (\ref{L-2-mPenner-M3}). In this case $\widetilde{\psi}_{\alpha}(x)$ is identified with the four-point function of fields with momenta $\alpha$ and $\alpha_i$ for $i=1,2,3$ that are parameters of the potential (\ref{m_penner_potential}). One can also analogously write down higher level BPZ equations.


\subsection{Quantization and the topological recursion}   \label{ssec-toprec}

We also briefly summarize the construction of quantum curves and wave-functions by means of the topological recursion, and in particular $\beta$-deformed topological recursion in case $\beta\neq 1$. The main advantage of this formulation is that it works more generally, for a large class of algebraic curves, which are not necessarily spectral curves of some matrix models. More details of this formulation, which are quite technical and involved, are presented in \cite{Manabe:2015kbj}.

Recall that for the $\beta$-deformed matrix model, the connected $h$-point differentials
\begin{equation}
W_h(x_1,\ldots,x_h)=\beta^{h/2}\bigg<\prod_{i=1}^h\sum_{a=1}^N\frac{dx_i}{x_i-z_a}\bigg>^{(\mathrm{c})}
\label{h_conn_diff-Vir}
\end{equation}
are defined through the connected part $\left<X\right>^{(\mathrm{c})}$ of the normalized expectation value
\begin{equation}
\frac{\left<X\right>}{Z}
=\frac{1}{(2\pi)^N N!Z}\int_{{\IR}^N}  \Delta(z)^{2\beta}X\, e^{-\frac{\sqrt{\beta}}{\hbar}\sum_{a=1}^NV(z_a)}  \prod_{a=1}^N dz_a.
\label{v_exp_d}
\end{equation}
In the 't Hooft limit $N\to\infty$, $\hbar\to 0$, $\mu=\beta^{1/2}\hbar N=const$, (\ref{h_conn_diff-Vir}) has an asymptotic expansion
\begin{equation}
W_h(x_1,\ldots,x_h)=\sum_{g,\ell=0}^{\infty}\hbar^{2g-2+h+\ell}\gamma^{\ell}W^{(g,h)}_{\ell}(x_1,\ldots,x_h),
\label{W_diff_asy}
\end{equation}
where $\gamma=\beta^{1/2}-\beta^{-1/2}$. The so-called multi-resolvents $W^{(g,h)}_{\ell}$ that arise in this expansion satisfy a set of recurrence equations referred to as the ($\beta$-deformed) topological recursion -- which are reformulation of the loop equations -- with the initial condition specified by the spectral curve \cite{Eynard:2008mz,Chekhov:2009mm,Chekhov:2010zg}. Similarly as in the original (unrefined) topological recursion \cite{eyn-or}, one can regard the recurrence equations as a definition of multi-resolvents assigned to a given algebraic curve, and not necessarily related to (the existence of) a matrix model.

Recall now that the wave-function (\ref{Psi-alpha}) can be regarded as an expectation value of a  determinant-like insertion. Taking advantage of the properties of the determinant, the asymptotic expansion of this wave-function can be shown to take form
\begin{align}
\begin{split}
\quad \log \frac{\widehat{\psi}_{\alpha}(x)}{Z}&
=
-\frac{2\alpha}{\epsilon_1 \epsilon_2}V(x)+
\sum_{h=1}^{\infty}\frac{1}{h!}\Big(-\frac{2\alpha}{g_s}\Big)^h \int^{x}_{\infty}\cdots\int^{x}_{\infty}W_h(x_1',\ldots,x_h') \\
&=
\sum_{g,\ell=0, h=1}^{\infty}  \frac{(-1)^{g+\ell+h-1}}{h!} 2^{2-2g-\ell} \alpha^h (\epsilon_1\epsilon_2)^{g-1}
(\epsilon_1+\epsilon_2)^{\ell}F^{(g,h)}_{\ell}(x,\ldots,x).
\label{det_op_exp}
\end{split}
\end{align}
This equation can be also regarded as a definition of the wave-function assigned to a given algebraic curve, even in case there is no underlying matrix model. One can then reconstruct the operator that annihilates this wave-function order by order. In this way, starting from a (classical) algebraic curve, the quantum curve (at appropriate level, determined by the value of $\alpha$) can be reconstructed.

One important remark concerning the resulting wave-function is that it depends on the choice of the base point of integrals. In (\ref{det_op_exp}) this point is taken to infinity, and it leads to quantum curves that coincide with those discussed in previous chapters (in case the relevant matrix model exists). Another choice, considered e.g. in \cite{Norbury-quantum,Dumitrescu:2015mpa} is to identify the base point with the conjugate point $\overline{x}$. In \cite{Manabe:2015kbj} it is shown that wave-functions corresponding to these two base points are related in some definite way, and this phenomenon is analyzed in more detail in \cite{Bouchard:2016obz}. More details and various calculations using the formalism of the ($\beta$-deformed) topological recursion are presented in \cite{Manabe:2015kbj}.


\section{Higher level (super-Virasoro) quantum curves}  \label{sec-super}

In this section we consider quantum curves which have the structure of singular vectors of super-Virasoro algebra. This algebra underlies the supersymmetric version of a matrix model, also known as super-eigenvalue model, which we therefore analyze. This analysis is quite analogous to what we presented in section \ref{sec-Virasoro}, and we follow the original presentation in \cite{Ciosmak:2016wpx}.


\subsection{$\alpha/\beta$-deformed super-eigenvalue models}

To start with we introduce the $\beta$-deformed super-eigenvalue model, which is a $\beta$-deformation of the super-eigenvalue model introduced in \cite{AlvarezGaume:1991jd} and analyzed in \cite{AlvarezGaume:1991jd,Becker:1992rk,Plefka:1996tt}. This super-eigenvalue model was formulated during the search of a supersymmetric generalization of a hermitian matrix model. Even though a formulation in terms of a proper integral over some supersymmetric ensemble of matrices has not been found, it was shown that the super-eigenvalue model satisfies constraint equations that form $\mathcal{N}=1$ super-Virasoro algebra, analogously to Virasoro constraints for hermitian matrix model. This super-Virasoro algebra is defined by the following (anti)commutation relations
\begin{align}
\begin{split}
\{G_r,G_s\}&=2L_{r+s}+\frac{c}{3}\big(r^2-\frac14\big)\delta_{r+s,0},\\
[L_m,G_r]&=\big(\frac{m}{2}-r\big)G_{m+r},\\
[L_m,L_n]&=(m-n)L_{m+n}+\frac{c}{12}(m^3-m)\delta_{m+n,0},
\label{s_vir_alg}
\end{split}
\end{align}
and of our interest is the NS (Neveu-Schwarz) sector, in which
the indices of generators $G_r$ take half-integer values, $r\in\mathbb{Z}+\frac{1}{2}.$
Generators $L_n$ form the Virasoro subalgebra of the NS superalgebra, and their indices $n$ are integer.

The partition function of the $\beta$-deformed super-eigenvalue model that we consider takes form of a formal integral over an even number $N$ of bosonic eigenvalues $z_a$ and fermionic eigenvalues $\vartheta_a$
\begin{equation}
Z= \int\prod_{a=1}^Ndz_ad\vartheta_a\Delta(z,\vartheta)^{\beta}e^{-\frac{\sqrt{\beta}}{\hbar}\sum_{a=1}^NV(z_a,\vartheta_a)},  
\label{matrix_def}
\end{equation}
where
\begin{equation}
\Delta(z,\vartheta)=\prod_{1\le a<b\le N}(z_a-z_b-\vartheta_a\vartheta_b),
\end{equation}
and the potential depends on bosonic $t_n$ and fermionic $\xi_{n+1/2}$ times
\begin{align}
\begin{split}
V(x,\theta)&=V_{B}(x)+V_{F}(x)\theta,\\
V_{B}(x)&=\sum_{n=0}^{\infty}t_nx^n,\ \ \ \
V_{F}(x)=\sum_{n=0}^{\infty}\xi_{n+1/2}x^n,
\end{split}
\end{align}
such that $\{\vartheta_a, \xi_{n+1/2}\}=0$. Instead of $\hbar$ and $\beta$, we also use another pair of parameters (note slight different normalization compared to (\ref{matparam-intro}))
\begin{equation}
\epsilon_1=-\beta^{1/2}\hbar,\qquad   \epsilon_2=\beta^{-1/2}\hbar.   \label{e1e2}
\end{equation}
In what follows by $\left<\cdots\right>$ we denote an unnormalized expectation value, defined as
\begin{equation}
\left<\mathcal{O}\right> = \int\prod_{a=1}^Ndz_ad\vartheta_a\Delta(z,\vartheta)^{\beta}e^{-\frac{\sqrt{\beta}}{\hbar}\sum_{a=1}^NV(z_a,\vartheta_a)}\mathcal{O}.
\label{def_unnorm_exp}
\end{equation}
In the analysis of the partition function (\ref{matrix_def}) or expectation values of time-independent operators, it is useful to identify their time derivatives with the following expectation values
\begin{equation}
\Big\langle \sum_{a=1}^Nz_a^n \cdots \Big\rangle = -\frac{\hbar}{\sqrt{\beta}}\partial_{t_{n}} \Big\langle\cdots \Big\rangle,\qquad
\Big\langle \sum_{a=1}^Nz_a^n\vartheta_a\cdots\Big\rangle =  -\frac{\hbar}{\sqrt{\beta}}\partial_{\xi_{n+1/2}}\Big\langle\cdots \Big\rangle,
\label{id_mat_free}
\end{equation}
analogously to (\ref{dt=z_a}). Similarly as in (\ref{phi-field}), we can associate to the super-eigenvalue model free boson and free fermion fields, and using the above identifications they can be written as
\begin{align}
\phi(x)&=\frac{1}{\hbar}V_{B}(x)-\sqrt{\beta}\sum_{a=1}^N\log(x-z_a),
\label{free_boson}
\\
\psi(x)&=\frac{1}{\hbar}V_{F}(x)-\sqrt{\beta}\sum_{a=1}^N\frac{\vartheta_a}{x-z_a}.
\label{free_fermion}
\end{align}
We introduce then a supersymmetric wave-function; it can be defined by means of the above fields (analogously to the non-supersymmetric case, as mentioned below (\ref{dt=z_a})), and we also call it as the $\alpha/\beta$-deformed super-eigenvalue integral
\begin{equation}
\widehat{\chi}_{\alpha}(x,\theta)=\left<e^{\frac{\alpha}{\hbar}\big(\phi(x) + \psi(x)\theta \big)}\right>,
\label{chi_hat_def}
\end{equation}
where $\theta$ is a fermionic variable with $\{\theta, \vartheta_a\}=\{\theta, \xi_{n+1/2}\}=0$. The momentum $\alpha$ is bosonic, and we find that finite order differential equations for $\widehat{\chi}_{\alpha}(x,\theta)$ arise only for some special values of $\alpha$.
We also decompose the wave-function into bosonic $\widehat{\chi}_{B,\alpha}(x)$ and fermionic $\widehat{\chi}_{F,\alpha}(x)$ components
\be
\begin{split}
\widehat{\chi}_{\alpha}(x,\theta) = &
\widehat{\chi}_{B,\alpha}(x)+\widehat{\chi}_{F,\alpha}(x)\theta,  \\
& \widehat{\chi}_{B,\alpha}(x)\equiv
\left<e^{\frac{\alpha}{\hbar}\phi(x)}\right>=\widehat{\chi}_{\alpha}(x,0), \\
& \widehat{\chi}_{F,\alpha}(x)\equiv
\frac{\alpha}{\hbar}\left<\psi(x)e^{\frac{\alpha}{\hbar}\phi(x)}\right>=-\partial_{\theta}\widehat{\chi}_{\alpha}(x,\theta).
\label{wave_bose_fermi}
\end{split}
\ee

For the super-eigenvalue model we find a representation of the super-Virasoro algebra analogous to (\ref{sp_hat_f_op-intro}), which will be used to build higher level quantum curves
\be
\begin{split}
\widehat{G}_{-1/2} &= \theta\partial_x-\partial_{\theta},   \qquad \qquad \qquad \widehat{L}_{-1} = \partial_x,    \\
\widehat{G}_{-n+1/2} &= \frac{1}{\hbar^2(n-2)!}\Big(
\partial_x^{n-2}\big(V_B'(x)V_F(x)\big)+Q\hbar\partial_x^{n-1}V_F(x)
+\partial_x^{n-2}\widehat{h}(x)\Big),\ \ \textrm{for $n\ge 2$},
\label{h_g_chi_rep}   \\ 
\widehat{L}_{-n} & =\frac{1}{\hbar^2(n-2)!}\Big(\frac12\partial_x^{n-2}\big(V_{B}'(x)\big)^2
+\frac12\partial_x^{n-2}\big(V_F'(x)V_F(x)\big)  +  \\
&\qquad \qquad
+\frac12Q\hbar\partial_x^nV_{B}(x)
+\partial_x^{n-2}\widehat{f}(x)\Big),\ \ \textrm{for $n\ge 2$}.
\end{split}
\ee
Here we denote
\be
\begin{split}
\widehat{h}(x) = & \widehat{h}_t(x)+\widehat{h}_{\xi}(x), \label{h_x_op} \\ 
& \widehat{h}_t(x)\equiv\hbar^2\sum_{n=0}^{\infty}x^n\sum_{k=n+1}^{\infty}\xi_{k+1/2}\partial_{t_{k-n-1}},\\
& \widehat{h}_{\xi}(x)\equiv\hbar^2\sum_{n=0}^{\infty}x^n\sum_{k=n+2}^{\infty}kt_{k}\partial_{\xi_{k-n-3/2}},
\end{split}
\ee
and
\be
\begin{split}
\widehat{f}(x)=&\widehat{f}_t(x)+\widehat{f}_{\xi}(x),
\label{f_x_op}  \\  
& \widehat{f}_t(x)\equiv\hbar^2\sum_{n=0}^{\infty}x^n\sum_{k=n+2}^{\infty}kt_{k}\partial_{t_{k-n-2}}, \\
& \widehat{f}_{\xi}(x)\equiv\hbar^2\sum_{n=0}^{\infty}x^n\sum_{k=n+2}^{\infty}\Big(k-\frac{n+1}{2}\Big)\xi_{k+1/2}\partial_{\xi_{k-n-3/2}},
\end{split}
\ee
and inductively we define
\begin{equation}
\partial_x^{n}\widehat{h}(x) \equiv \left[\partial_x, \partial_x^{n-1}\widehat{h}(x)\right],\qquad\quad
\partial_x^{n}\widehat{f}(x) \equiv
\left[\partial_x, \partial_x^{n-1}\widehat{f}(x)\right].
\end{equation}
By direct calculation one can check  that the generators in (\ref{h_g_chi_rep}) indeed satisfy the super-Virasoro algebra, in particular
$\big\{\widehat{G}_{-m-1/2},\widehat{G}_{-n-1/2}\big\}=2\widehat{L}_{-m-n-1}$, $\big[\widehat{L}_{-m},\widehat{G}_{-n-1/2}\big]=\big(n-(m-1)/2\big)\widehat{G}_{-m-n-1/2}$ and $\big[\widehat{L}_{-m},\widehat{L}_{-n}\big]=(n-m)\widehat{L}_{-m-n}$ for $n\ge 1$.

Using (\ref{h_g_chi_rep}) we also introduce another representation of the super-Virasoro algebra, as acting on bosonic and fermionic components $\widehat{\chi}_{B,\alpha}(x)$ and $\widehat{\chi}_{F,\alpha}(x)$ in (\ref{wave_bose_fermi});
for $n\ge 2$ we get
\begin{align}
\begin{split}
&
\widehat{\mathsf{G}}_{-1/2}\widehat{\chi}_{B,\alpha}(x)=
\widehat{\chi}_{F,\alpha}(x),\qquad
\widehat{\mathsf{G}}_{-1/2}\widehat{\chi}_{F,\alpha}(x)=
\partial_x\widehat{\chi}_{B,\alpha}(x),\\
&
\widehat{\mathsf{G}}_{-n+1/2}\widehat{\chi}_{B,\alpha}(x)=
\widehat{G}_{-n+1/2}\widehat{\chi}_{B,\alpha}(x),\qquad
\widehat{\mathsf{G}}_{-n+1/2}\widehat{\chi}_{F,\alpha}(x)=
\widehat{G}_{-n+1/2}\widehat{\chi}_{F,\alpha}(x),
\label{h_g_bf_chi_rep} \\ 
&
\widehat{\mathsf{L}}_{-1}\widehat{\chi}_{B,\alpha}(x)=\partial_x\widehat{\chi}_{B,\alpha}(x),\qquad
\widehat{\mathsf{L}}_{-1}\widehat{\chi}_{F,\alpha}(x)=\partial_x\widehat{\chi}_{F,\alpha}(x),\\
&
\widehat{\mathsf{L}}_{-n}\widehat{\chi}_{B,\alpha}(x)=
\widehat{L}_{-n}\widehat{\chi}_{B,\alpha}(x),\qquad
\widehat{\mathsf{L}}_{-n}\widehat{\chi}_{F,\alpha}(x)=
\widehat{L}_{-n}\widehat{\chi}_{F,\alpha}(x).
\end{split}
\end{align}
From this representation, $\widehat{\mathsf{G}}_{-1/2}^2=\widehat{\mathsf{L}}_{-1}$ on $\widehat{\chi}_{B,\alpha}(x)$, and we find commutation relations
$\big\{\widehat{\mathsf{G}}_{-m-1/2},\widehat{\mathsf{G}}_{-n-1/2}\big\}=2\widehat{\mathsf{L}}_{-m-n-1}$, $\big[\widehat{\mathsf{L}}_{-m},\widehat{\mathsf{G}}_{-n-1/2}\big]=\big(n-(m-1)/2\big)\widehat{\mathsf{G}}_{-m-n-1/2}$, as well as $\big[\widehat{\mathsf{L}}_{-m},\widehat{\mathsf{L}}_{-n}\big]=(n-m)\widehat{\mathsf{L}}_{-m-n}$ for $n\ge 1$. Although commutation relations
\begin{equation}
\big\{\widehat{\mathsf{G}}_{-1/2},\widehat{\mathsf{G}}_{-n+1/2}\big\}=2\widehat{\mathsf{L}}_{-n},\qquad
\big[\widehat{\mathsf{G}}_{-1/2},\widehat{\mathsf{L}}_{-n}\big]=\frac{n-1}{2}\widehat{\mathsf{G}}_{-n-1/2},\qquad n\ge 1,
\end{equation}
are not obvious at this stage, they are needed to construct super-quantum curves.

Yet another representation of super-Virasoro algebra that we consider is associated to the wave-function normalized by (\ref{matrix_def})
\begin{equation}
\Psi_{\alpha}(x,\theta) = \frac{\widehat{\chi}_{\alpha}(x,\theta)}{Z}.   \label{Psi-def}
\end{equation}
In this case the operators $\widehat{G}_{n+1/2}$ and $\widehat{L}_n$ in (\ref{h_g_chi_rep}) are converted to $\widehat{\mathcal{G}}_{-n+1/2}\equiv
Z^{-1}\widehat{G}_{-n+1/2}Z$ and $\widehat{\mathcal{L}}_{-n}\equiv
Z^{-1}\widehat{L}_{-n}Z$, and for $n\geq 2$ they take form
\be
\begin{split}
\widehat{\mathcal{G}}_{-n+1/2} & =
\frac{1}{\hbar^2(n-2)!}\Big(
\partial_x^{n-2}\big(V_B'(x)V_F(x)\big)+Q\hbar\partial_x^{n-1}V_F(x)
+\partial_x^{n-2}\widehat{h}(x) +
\\
&\qquad +\big[\partial_x^{n-2}\widehat{h}(x),\log Z\big]\Big),   \\
\widehat{\mathcal{L}}_{-n} &=
\frac{1}{\hbar^2(n-2)!}\Big(
\frac12\partial_x^{n-2}\big(V_{B}'(x)^2\big)
+\frac12\partial_x^{n-2}\big(V_F'(x)V_F(x)\big)
+\frac12Q\hbar\partial_x^nV_{B}(x)  + \\
&\qquad
+\partial_x^{n-2}\widehat{f}(x)
+\big[\partial_x^{n-2}\widehat{f}(x),\log Z\big]\Big).   \label{Psi-GL}
\end{split}
\ee


\subsection{Higher level quantum curves as super-Virasoro singular vectors}

We present now super-quantum curves $\widehat{A}_{n}^{\alpha}$, which annihilate wave-functions  defined in (\ref{chi_hat_def})
\begin{equation}
\widehat{A}_{n}^{\alpha}\widehat{\chi}_{\alpha}(x,\theta)=0.   \label{Achi0}
\end{equation}
These super-quantum curves have the structure of the Neveu-Schwarz (NS) super-Virasoro singular vectors. Recall that NS super-Virasoro singular vectors exist only at levels $n=pq/2$, for integer $p$ and $q$ such that $p-q$ is even, and correspond to degenerate momenta of the form
\begin{equation}
\alpha=\alpha_{p,q} = \frac{(p-1)\beta^{1/2}-(q-1)\beta^{-1/2}}{2}\hbar,
\qquad \textrm{with $p-q\in 2{\IZ}$}  \label{s_vir_sing_momenta}
\end{equation}
which label primary states $|\Delta_{p,q}\rangle$ of weight $\Delta_{p,q}=-\frac{pq-1}{4} + \frac{p^2-1}{8}\beta + \frac{q^2-1}{8}\frac{1}{\beta}$. The first few NS super-Virasoro singular vectors take form
\begin{align}
\textrm{level}\ \frac12:&\qquad G_{-1/2}|\Delta_{1,1}\rangle  \nonumber \\
\textrm{level}\ \frac32:&\qquad  \left(L_{-1}G_{-1/2} - \beta G_{-3/2}\right)|\Delta_{3,1}\rangle  \label{NS-singular} \\
\textrm{level}\ 2:&\qquad   \Big(L_{-1}^2 -\frac12\big(\beta^{-1/2} - \beta^{1/2}\big)^2 L_{-2} - G_{-3/2}G_{-1/2}\Big)|\Delta_{2,2}\rangle  \nonumber \\
\textrm{level}\ \frac52:&\qquad   \left(L_{-1}^2 G_{-1/2} +2\beta(3\beta -1) -3\beta G_{-3/2}L_{-1} -2\beta L_{-2} G_{-1/2}\right)|\Delta_{5,1}\rangle \nonumber
\end{align}

Imposing the condition that the wave-function $\widehat{\chi}_{\alpha}(x,\theta)$ satisfies a differential equation of a finite order in $x$, we find that it can be satisfied only for special values of $\alpha$ in (\ref{s_vir_sing_momenta}), and operators $\widehat{A}^n_{\alpha}$ encode the structure of NS singular vectors, such as those in (\ref{NS-singular}). In the notation $\widehat{A}_{n}^{\alpha}$ the superscript $\alpha$ refers to the deformation parameter of the corresponding wave-function, and the subscript $n$ denotes the level of the singular vector encoding the structure of a given super-quantum curve. In particular the super-quantum curve at level 3/2 reduces to super-spectral curve of the super-eigenvalue model in the classical limit.

The operators $\widehat{A}_{n}^{\alpha}$ encode the structure of super-Virasoro singular vectors as follows. First, using the decomposition (\ref{wave_bose_fermi}) and the representation (\ref{h_g_bf_chi_rep}), we write the wave-function as
\be
\widehat{\chi}_{\alpha}(x,\theta) = (1-\theta  \widehat{\mathsf{G}}_{-1/2} ) \widehat{\chi}_{B,\alpha}(x),
\ee
so that all information is essentially encoded in its bosonic component $\widehat{\chi}_{B,\alpha}(x)$. Note that
\be
\theta \widehat{\chi}_{\alpha}(x,\theta)  = \theta \widehat{\chi}_{B,\alpha}(x), \qquad \widehat{\chi}_{F,\alpha}(x) = -\partial_{\theta} \widehat{\chi}_{\alpha}(x,\theta)  = \widehat{\mathsf{G}}_{-1/2}  \widehat{\chi}_{B,\alpha}(x).     \label{theta-G12}
\ee
In consequence one can introduce an operator $\widehat{\mathsf{A}}_{n}^{(0)}$ which annihilates the bosonic wave-function
\be
\widehat{\mathsf{A}}_{n}^{(0)}\widehat{\chi}_{B,\alpha_{p,q}}(x) = 0.  \label{A0-chiB}
\ee
As we will show, this is the operator $\widehat{\mathsf{A}}_{n}^{(0)}$ that takes form of a super-Virasoro singular vector (\ref{NS-singular}), written in terms of generators (\ref{h_g_bf_chi_rep}). The above equation can be also rewritten in terms of an operator $\widehat{A}_{n}^{(0)}$ -- which arises by replacing $\widehat{\mathsf{G}}_{-1/2}$ by $(-\partial_{\theta})$ in $\widehat{\mathsf{A}}_{n}^{(0)}$ -- acting on $\widehat{\chi}_{\alpha}(x,\theta)$
\begin{equation}
\theta\widehat{A}_{n}^{(0)}\widehat{\chi}_{\alpha_{p,q}}(x,\theta)=
\theta\widehat{\mathsf{A}}_{n}^{(0)}\widehat{\chi}_{B,\alpha_{p,q}}(x)=0.
\label{hq_curve_gen_c}
\end{equation}
Furthermore, the bosonic wave-function is annihilated also by the following operator $\widehat{\mathsf{A}}_{n}^{(1)}$ (expressed in terms of generators (\ref{h_g_bf_chi_rep}))
\begin{equation}
\widehat{\mathsf{A}}_{n}^{(1)}=
\widehat{\mathsf{G}}_{-1/2}\widehat{\mathsf{A}}_{n}^{(0)},
\end{equation}
related to the operator acting on $\widehat{\chi}_{\alpha}(x,\theta)$ as
\begin{equation}
\theta\widehat{A}_{n}^{(1)}\widehat{\chi}_{\alpha_{p,q}}(x,\theta)=
\theta\widehat{\mathsf{A}}_{n}^{(1)}\widehat{\chi}_{B,\alpha_{p,q}}(x)=0.
\end{equation}
The original super-quantum curve $\widehat{A}_{n}^{\alpha}$ can be reconstructed from $\widehat{A}_{n}^{(0)}$ and $\widehat{A}_{n}^{(1)}$ given above
\begin{equation}
\widehat{A}_{n}^{\alpha}=
\widehat{A}_{n}^{(0)}-\theta\partial_{\theta}\widehat{A}_{n}^{(0)}
-\theta\widehat{A}_{n}^{(1)},
\label{q_curve_gen_c}
\end{equation}
and this operator is expressed in terms of generators (\ref{h_g_chi_rep}). The operator (\ref{q_curve_gen_c}) can be also transformed into an operator that annihilates the normalized wave-function (\ref{Psi-def})
\begin{equation}
\widehat{\mathcal{A}}_n^{\alpha} = Z^{-1}\widehat{A}_n^{\alpha}Z,   \label{A-hat-norm}
\end{equation}
which is then expressed in terms of generators (\ref{Psi-GL}).

We present now explicit form of super-quantum curves at first few levels. At level 3/2,
for specific values of momenta
\begin{equation}
\alpha=0,\quad \beta^{1/2}\hbar,\quad \textrm{or}\quad -\beta^{-1/2}\hbar,
\label{momenta_3_2}
\end{equation}
we find
\begin{equation}
\widehat{A}_{3/2}^{(0)}=-\partial_x\partial_{\theta}
-\frac{\alpha^2}{\hbar^2}\widehat{G}_{-3/2},
\end{equation}
and as the operators acting on $\widehat{\chi}_{B}(x)$
\begin{align}
\begin{split}
&
\widehat{\mathsf{A}}_{3/2}^{(0)}=
\widehat{\mathsf{L}}_{-1}\widehat{\mathsf{G}}_{-1/2}-\frac{\alpha^2}{\hbar^2}\widehat{\mathsf{G}}_{-3/2},
\\
&
\widehat{\mathsf{A}}_{3/2}^{(1)}=\widehat{\mathsf{G}}_{-1/2}\widehat{\mathsf{A}}_{3/2}^{(0)}=
\widehat{\mathsf{L}}_{-1}^2-\frac{2\alpha^2}{\hbar^2}\widehat{\mathsf{L}}_{-2}
+\frac{\alpha^2}{\hbar^2}\widehat{\mathsf{G}}_{-3/2}\widehat{\mathsf{G}}_{-1/2}.
\label{q_curve_eq_c_3_2}
\end{split}
\end{align}
We find that $\widehat{\mathsf{A}}_{3/2}^{(0)}$ indeed has the form of singular vectors at level $3/2$, see (\ref{NS-singular}), for $\alpha=\pm \beta^{\pm 1/2} \hbar$. For the remaining value $\alpha=0$ in (\ref{momenta_3_2}), $\widehat{\mathsf{A}}_{3/2}^{(0)}$ reduces to the singular vector at level $1/2$, see (\ref{NS-singular}), which can be identified with the partition function (\ref{matrix_def}). We also obtain
\begin{equation}
\widehat{A}_{3/2}^{(1)}=
\partial_x^2-\frac{2\alpha^2}{\hbar^2}\widehat{L}_{-2}
-\frac{\alpha^2}{\hbar^2}\widehat{G}_{-3/2}\partial_{\theta},
\end{equation}
and from (\ref{q_curve_gen_c}) we finally find the super-quantum curve equation
\begin{equation}
\widehat{A}_{3/2}^{\alpha}\widehat{\chi}_{\alpha}(x,\theta)=0,\qquad
\widehat{A}_{3/2}^{\alpha}=
-\partial_x\partial_{\theta}-\frac{\alpha^2}{\hbar^2}\widehat{G}_{-3/2}
-\theta\Big(\partial_x^2-\frac{2\alpha^2}{\hbar^2}\widehat{L}_{-2}\Big),
\label{q_curve_eq_3_2}
\end{equation}
for special values of $\alpha$ in (\ref{momenta_3_2}).

At level 2,
for specific values of momenta
\begin{equation}
\alpha=0,\quad \beta^{1/2}\hbar, \quad -\beta^{-1/2}\hbar, \quad -\frac{Q\hbar}{2} =\frac{\beta^{1/2}-\beta^{-1/2}}{2}\hbar,
\label{momenta_2}
\end{equation}
we find
\begin{equation}
\widehat{A}_{2}^{(0)}=
\partial_x^2-\frac{2\alpha^2}{\hbar^2}\widehat{L}_{-2}
-\frac{2\alpha^2+Q\hbar\alpha-\hbar^2}{\hbar^2}\widehat{G}_{-3/2}\partial_{\theta}.
\end{equation}
This leads to the following operators acting on $\widehat{\chi}_{B,\alpha}(x)$
\begin{align}
\begin{split}
\widehat{\mathsf{A}}_{2}^{(0)}&=
\widehat{\mathsf{L}}_{-1}^2-\frac{2\alpha^2}{\hbar^2}\widehat{\mathsf{L}}_{-2}
+\frac{2\alpha^2+Q\hbar\alpha-\hbar^2}{\hbar^2}\widehat{\mathsf{G}}_{-3/2}\widehat{\mathsf{G}}_{-1/2} =  \\
&=\widehat{\mathsf{A}}_{3/2}^{(1)}
+\frac{\alpha^2+Q\hbar\alpha-\hbar^2}{\hbar^2}\widehat{\mathsf{G}}_{-3/2}\widehat{\mathsf{G}}_{-1/2},  \\
\widehat{\mathsf{A}}_{2}^{(1)}&=\widehat{\mathsf{G}}_{-1/2}\widehat{\mathsf{A}}_{2}^{(0)}  = \widehat{\mathsf{L}}_{-1}^2\widehat{\mathsf{G}}_{-1/2}
-\frac{2\alpha^2}{\hbar^2}\widehat{\mathsf{L}}_{-2}\widehat{\mathsf{G}}_{-1/2}
-\frac{\alpha^2}{\hbar^2}\widehat{\mathsf{G}}_{-5/2} + \\
&\qquad \qquad \qquad \quad
+\frac{2\alpha^2+Q\hbar\alpha-\hbar^2}{\hbar^2}\big(2\widehat{\mathsf{L}}_{-2}\widehat{\mathsf{G}}_{-1/2}-\widehat{\mathsf{G}}_{-3/2}\widehat{\mathsf{L}}_{-1}\big),
\label{q_curve_eq_c_2}
\end{split}
\end{align}
and for $\alpha=-\frac{Q\hbar}{2}$ the operator $\widehat{\mathsf{A}}_{2}^{(0)}$ takes form of singular vector at level 2 (\ref{NS-singular}). Furthermore
\begin{equation}
\widehat{A}_{2}^{(1)}=
-\partial_x^2\partial_{\theta}
+\frac{2\alpha^2}{\hbar^2}\widehat{L}_{-2}\partial_{\theta}
-\frac{\alpha^2}{\hbar^2}\widehat{G}_{-5/2}
-\frac{2\alpha^2+Q\hbar\alpha-\hbar^2}{\hbar^2}\big(2\widehat{L}_{-2}\partial_{\theta}+\widehat{G}_{-3/2}\partial_x\big),
\end{equation}
and finally from (\ref{q_curve_gen_c}) we obtain the super-quantum curve equation at level 2
\begin{align}
\begin{split}
&
\widehat{A}_{2}^{\alpha}\widehat{\chi}_{\alpha}(x,\theta) = 0,   \\
&
\widehat{A}_{2}^{\alpha} =
\partial_x^2-\frac{2\alpha^2}{\hbar^2}\widehat{L}_{-2}
-\frac{2\alpha^2+Q\hbar\alpha-\hbar^2}{\hbar^2}\widehat{G}_{-3/2}\partial_{\theta}
+\frac{\alpha^2}{\hbar^2}\theta\widehat{G}_{-5/2}  +  \\
&\hspace{2.5em}
+\frac{2\alpha^2+Q\hbar\alpha-\hbar^2}{\hbar^2}\theta
\Big(2\widehat{L}_{-2}\partial_{\theta}+\widehat{G}_{-3/2}\partial_x\Big),
\label{q_curve_eq_2}
\end{split}
\end{align}
for the momenta (\ref{momenta_2}). Upon substitution those values of $\alpha$ that correspond to lower levels (\ref{momenta_3_2}), this result reduces to quantum curves at lower levels.

At level 5/2,
for specific values of momenta
\begin{equation}
\alpha=0,\quad \beta^{1/2}\hbar, \quad -\beta^{-1/2}\hbar, \quad -\frac{Q\hbar}{2},\quad 2\beta^{1/2}\hbar, \quad -2\beta^{-1/2}\hbar,
\label{momenta_5_2}
\end{equation}
we find
\begin{align}
\begin{split}
\widehat{A}_{5/2}^{(0)}&=
-\partial_x^2\partial_{\theta}
+\frac{2\alpha(\alpha^2+Q\hbar\alpha-\hbar^2)}{\hbar^2(3\alpha+2Q\hbar)}\widehat{L}_{-2}\partial_{\theta}
-\frac{\alpha(2\alpha^2+Q\hbar\alpha+\hbar^2)}{\hbar^2(3\alpha+2Q\hbar)}\widehat{G}_{-3/2}\partial_x +
\\
&\ \ \
+\frac{\alpha^2(2\alpha^3+3Q\hbar\alpha^2+(Q^2-5)\hbar^2\alpha-3Q\hbar^3)}{\hbar^4(3\alpha+2Q\hbar)}\widehat{G}_{-5/2},
\end{split}
\end{align}
and acting on the bosonic component this operator is represented as
\begin{equation}
\widehat{\mathsf{A}}_{5/2}^{(0)}=\widehat{\mathsf{A}}_{2}^{(1)}
+\frac{(\alpha^2+Q\hbar\alpha-\hbar^2)(2\alpha+Q\hbar)}{\hbar^2(3\alpha+2Q\hbar)}
\Big(\frac{\alpha^2}{\hbar^2}\widehat{\mathsf{G}}_{-5/2}
+2\widehat{\mathsf{G}}_{-3/2}\widehat{\mathsf{L}}_{-1}
-4\widehat{\mathsf{L}}_{-2}\widehat{\mathsf{G}}_{-1/2}\Big).
\label{q_curve_eq_c_5_2}
\end{equation}
As expected, for $\alpha=\pm 2\beta^{\pm 1/2}\hbar$ it has the same form as a singular vector at level $5/2$ (\ref{NS-singular}). Finally, determining $\widehat{\mathsf{A}}_{5/2}^{(1)}$ and using (\ref{q_curve_gen_c}), we obtain the super-quantum curve at level $5/2$
\begin{align}
\begin{split}
&
\widehat{A}_{5/2}^{\alpha}\widehat{\chi}_{\alpha}(x,\theta)=0,
\\
&
\widehat{A}_{5/2}^{\alpha}=
-\partial_x^2\partial_{\theta}
+\frac{2\alpha(\alpha^2+Q\hbar\alpha-\hbar^2)}{\hbar^2(3\alpha+2Q\hbar)}\widehat{L}_{-2}\partial_{\theta}
-\frac{\alpha(2\alpha^2+Q\hbar\alpha+\hbar^2)}{\hbar^2(3\alpha+2Q\hbar)}\widehat{G}_{-3/2}\partial_x +
\\
&\hspace{2.5em}
+\frac{\alpha^2(2\alpha^3+3Q\hbar\alpha^2+(Q^2-5)\hbar^2\alpha-3Q\hbar^3)}{\hbar^4(3\alpha+2Q\hbar)}\widehat{G}_{-5/2}
-\theta\Big(\partial_x^3
-\frac{2\alpha^2}{\hbar^2}\widehat{L}_{-2}\partial_x +
\\
&\hspace{2.5em}
+\frac{\alpha(\alpha^2+Q\hbar\alpha-\hbar^2)}{\hbar^2(3\alpha+2Q\hbar)}\widehat{G}_{-5/2}\partial_{\theta}
+\frac{2\alpha^2(2\alpha^3+3Q\hbar\alpha^2+(Q^2-5)\hbar^2\alpha-3Q\hbar^3)}{\hbar^4(3\alpha+2Q\hbar)}\widehat{L}_{-3}\Big),
\label{q_curve_eq_5_2}
\end{split}
\end{align}
for the momenta (\ref{momenta_5_2}). For the momenta (\ref{momenta_2}) at level 2 the operator $\widehat{\mathsf{A}}_{5/2}^{(0)}$ reduces to the operator $\widehat{\mathsf{A}}_{2}^{(1)}$ in (\ref{q_curve_eq_c_2}), and encodes singular vectors up to level 2.


\subsection{Double quantum structure}

Similarly as in the bosonic case discussed in section \ref{ssec-double}, also two classical limits and correspondingly a double quantum structure can be considered in the supersymmetric case. The first classical limit is the large $N$ 't Hooft limit (of vanishing $\hbar$), and the second limit is that of infinite central charge in the conformal field theory interpretation.

Let us consider 't Hooft limit first, such that $N\to \infty$, $\hbar\to 0$, and $\mu=\beta^{1/2} \hbar N=const$. In this limit we introduce
\begin{equation}
y_B(x) = \lim_{\begin{subarray}{c}N\to\infty\\\beta=1 \end{subarray}}\frac{\hbar}{Z}\big<\partial_x\phi(x)\big>,\qquad
y_F(x) = \lim_{\begin{subarray}{c}N\to\infty\\\beta=1 \end{subarray}}\frac{\hbar}{Z}\big<\psi(x)\big>,
\end{equation}
and then we find
\begin{align}
\begin{split}
&
\hbar\partial_{\theta}\Psi_{\alpha}(x,\theta)\
\stackrel{N\to\infty, \ \beta\to 1}{\longrightarrow}\
-\frac{\alpha}{\hbar}y_F(x)\Psi_{\alpha}(x,\theta),
\\
&
\hbar\partial_x\Psi_{\alpha}(x,\theta)\
\stackrel{N\to\infty, \ \beta\to 1}{\longrightarrow}\
\frac{\alpha}{\hbar}\big(y_B(x)+y_F'(x)\theta\big)
\Psi_{\alpha}(x,\theta).
\label{par_h_classic}
\end{split}
\end{align}
A redefinition of the super-quantum curve at level $3/2$ in (\ref{q_curve_eq_3_2}), which annihilates the normalized wave-function (\ref{Psi-def}), takes form
\begin{equation}
-\hbar^2\widehat{\mathcal{A}}_{3/2}^{\alpha}=\hbar^2\partial_x\partial_{\theta}
+\alpha^2\widehat{\mathcal{G}}_{-3/2}
+\hbar^2\theta \partial_x^2
-2\theta\alpha^2\widehat{\mathcal{L}}_{-2},
\end{equation}
for momenta $\alpha=\pm \beta^{\pm 1/2}\hbar$. This quantum curve reduces to the supersymmetric algebraic curve, which can be written as
\begin{equation}
\widehat{\mathcal{A}}_{3/2}^{\alpha}\Psi_{\alpha}(x,\theta)=0\
\stackrel{N\to\infty, \ \beta\to 1}{\longrightarrow}\
A_F(x,y_B|y_F)=\theta A_B(x,y_B|y_F),  \label{super-spectral}
\end{equation}
where
\begin{align}
\begin{cases}
& A_F(x,y_B|y_F)\equiv y_B(x)y_F(x)+G(x)=0,
\\
& A_B(x,y_B|y_F)\equiv y_B(x)^2+y_F'(x)y_F(x)+2L(x)=0,
\label{b_f_sp_curve}
\end{cases}
\end{align}
and
\begin{equation}
G(x)=-\lim_{\begin{subarray}{c}\hbar\to 0\\\beta=1\end{subarray}}
\hbar^2\widehat{\mathcal{G}}_{-3/2},\qquad
L(x)=-\lim_{\begin{subarray}{c}\hbar\to 0\\\beta=1\end{subarray}}
\hbar^2\widehat{\mathcal{L}}_{-2}.   \label{GL-classical}
\end{equation}
The supersymmetric algebraic curve defined by (\ref{super-spectral}) coincides with the spectral curve of the super-eigenvalue model, which can be determined by the analysis of eigenvalue distribution in the large $N$ limit, and that we also refer to as super-spectral curve \cite{Ciosmak:2016wpx}. It follows that the super-quantum curve at level $3/2$ can be regarded as a quantization of the super-spectral curve. Classical curves at higher levels can be obtained as appropriate limits of the quantum curves in the second interesting limit (Nekrasov-Shatashvili limit) listed in (\ref{high_ns_q_ex}); for completeness, these classical curves at a few higher levels take form
\begin{align}
\begin{split}
\widehat{\mathcal{A}}_{5/2}^{\textrm{cl}}&=
-2\Big(3y_B(x)A_F(x,y_B|y_F)+2y_F(x)A_B(x,y_B|y_F)\Big) +
\\
&\ \
+2\theta\Big(3y_F'(x)A_F(x,y_B|y_F)+\partial_xA_F(x,y_B|y_F)y_F(x)
+8y_B(x)A_B(x,y_B|y_F)\Big),
\\
\widehat{\mathcal{A}}_{7/2}^{\textrm{cl}}&=
-18\Big(\big(3y_B(x)^2+L(x)\big)A_F(x,y_B|y_F)+3y_F(x)y_B(x)A_B(x,y_B|y_F)\Big) +
\\
&\ \
+9\theta\Big(2\big(9y_B(x)^2+3y_F'(x)y_F(x)+2L(x)\big)A_B(x,y_B|y_F) +
\\
&\ \
+\big(15y_F'(x)y_B(x)+3y_F(x)y_B'(x)+4\partial_xG(x)\big)A_F(x,y_B|y_F) +
\\
&\ \
+3G(x)\partial_xA_F(x,y_B|y_F)\Big).
\end{split}
\end{align}

The second limit of interests, that can be identified as the Nekrasov-Shatashvili limit or a classical limit in the super-Liouville theory, in terms of parameters (\ref{e1e2}) arises for $\epsilon_1\to 0$ with $\epsilon_2$ fixed. We consider normalized wave-functions $\Psi_{\alpha_{2p+1,1}}(x,\theta)$ for the momenta
\begin{equation}
\alpha=\alpha_{2p+1,1}=-p\epsilon_1,   \label{alpha-2p1}
\end{equation}
which in this limit factorize as
\begin{equation}
\Psi_{-p\epsilon_1}^{\textrm{NS}}(x,\theta)\equiv
\lim_{\epsilon_1\to 0}\Psi_{-p\epsilon_1}(x,\theta)=
\left(\Psi_{-\epsilon_1}^{\textrm{NS}}(x,\theta)\right)^p.
\label{wave_ns_factor}
\end{equation}
The corresponding super-quantum curves arise at level $p+1/2$, and we rescale them as follows
\begin{equation}
\widehat{\mathcal{A}}_{p+1/2}^{\textrm{NS}}\Psi_{-p\epsilon_1}^{\textrm{NS}}(x,\theta)=0,\qquad
\widehat{\mathcal{A}}_{p+1/2}^{\textrm{NS}} =
-\epsilon_2^{p+1}\lim_{\epsilon_1\to 0}
\widehat{\mathcal{A}}_{p+1/2}^{-p\epsilon_1},
\end{equation}
and express in terms of operators
\be
\begin{split}
\widehat{\mathcal{G}}_{-n+1/2}^{\textrm{NS}} & = \lim_{\epsilon_1\to 0}\epsilon_1\epsilon_2\widehat{\mathcal{G}}_{-n+1/2} = \\
&=-\frac{1}{(n-2)!}\Big(
\partial_x^{n-2}\big(V_B'(x)V_F(x)\big)+\epsilon_2\partial_x^{n-1}V_F(x)
+\partial_x^{n-2}F_{F}^{(0)}(x,\epsilon_2)\Big),    \\
\widehat{\mathcal{L}}_{-n}^{\textrm{NS}} & = \lim_{\epsilon_1\to 0}\epsilon_1\epsilon_2\widehat{\mathcal{L}}_{-n}= \\
& =-\frac{1}{(n-2)!}\Big(
\frac12\partial_x^{n-2}\big(V_{B}'(x)^2\big)
+\frac12\partial_x^{n-2}\big(V_F'(x)V_F(x)\big)
+\frac12\epsilon_2\partial_x^nV_{B}(x)
+\partial_x^{n-2}F_{B}^{(0)}(x,\epsilon_2)\Big),
\end{split}
\ee
where
\begin{equation}
F_{F}^{(0)}(x,\epsilon_2) =
-\frac{1}{\epsilon_1\epsilon_2}\big[\widehat{h}(x), F^{(0)}(\epsilon_2)\big],
\qquad
F_{B}^{(0)}(x,\epsilon_2) =
-\frac{1}{\epsilon_1\epsilon_2}\big[\widehat{f}(x), F^{(0)}(\epsilon_2)\big],
\end{equation}
are defined in terms of the deformed prepotential
\begin{equation}
F^{(0)}(\epsilon_2) =
-\lim_{\epsilon_1\to 0}\epsilon_1\epsilon_2\log Z.
\end{equation}
Quantities $\widehat{\mathcal{G}}_{-n+1/2}^{\textrm{NS}}$ and $\widehat{\mathcal{L}}_{-n}^{\textrm{NS}}$ are simply fermionic and bosonic functions of $x$, such that
\be
\partial_x^n\widehat{\mathcal{G}}_{-3/2}^{\textrm{NS}}=n!\widehat{\mathcal{G}}_{-n-3/2}^{\textrm{NS}}, \qquad \partial_x^n\widehat{\mathcal{L}}_{-2}^{\textrm{NS}}=n!\widehat{\mathcal{L}}_{-n-2}^{\textrm{NS}}.   \label{GL-NS-relations}
\ee

The super-quantum curve equation (\ref{q_curve_eq_3_2}) at level $3/2$ in the limit $\epsilon_1\to 0$ yields
\begin{equation}
\widehat{\mathcal{A}}_{3/2}^{\textrm{NS}}\Psi_{-\epsilon_1}^{\textrm{NS}}(x,\theta)=
\left(\epsilon_2^2\partial_x\partial_{\theta}-\widehat{\mathcal{G}}_{-3/2}^{\textrm{NS}}
+\theta\epsilon_2^2\partial_x^2
+2\theta\widehat{\mathcal{L}}_{-2}^{\textrm{NS}}\right)
\Psi_{-\epsilon_1}^{\textrm{NS}}(x,\theta)=0,
\label{q_curve_ns_3_2a}
\end{equation}
and equivalently can be written as
\begin{equation}
\theta\left(\epsilon_2^2\partial_x\partial_{\theta}-\widehat{\mathcal{G}}_{-3/2}^{\textrm{NS}}\right)
\Psi_{-\epsilon_1}^{\textrm{NS}}(x,\theta)=
\theta\left(\epsilon_2^2\partial_x^2+\widehat{\mathcal{G}}_{-3/2}^{\textrm{NS}}\partial_{\theta}
+2\widehat{\mathcal{L}}_{-2}^{\textrm{NS}}\right)
\Psi_{-\epsilon_1}^{\textrm{NS}}(x,\theta)=0.
\label{q_curve_ns_3_2}
\end{equation}
Higher level quantum curves in this limit can be determined by generalizing the recurrence relation (\ref{NS_B_op}) to the supersymmetric case \cite{Ciosmak:2016wpx}, which yields the following results
\begin{align}
\begin{split}
\widehat{\mathcal{A}}_{5/2}^{\textrm{NS}}&=
\epsilon_2^3\partial_x^2\partial_{\theta}
+2\epsilon_2\widehat{\mathcal{L}}_{-2}^{\textrm{NS}}\partial_{\theta}
-3\epsilon_2\widehat{\mathcal{G}}_{-3/2}^{\textrm{NS}}\partial_x
-2\epsilon_2\widehat{\mathcal{G}}_{-5/2}^{\textrm{NS}} +
\\
&\quad
+\theta\Big(\epsilon_2^3\partial_x^3
-\epsilon_2\widehat{\mathcal{G}}_{-5/2}^{\textrm{NS}}\partial_{\theta}
+8\epsilon_2\widehat{\mathcal{L}}_{-2}^{\textrm{NS}}\partial_x
+4\epsilon_2\widehat{\mathcal{L}}_{-3}^{\textrm{NS}}\Big),
\\
\widehat{\mathcal{A}}_{7/2}^{\textrm{NS}}&=
\epsilon_2^4\partial_x^3\partial_{\theta}
+8\epsilon_2^2\widehat{\mathcal{L}}_{-2}^{\textrm{NS}}\partial_x\partial_{\theta}
+4\epsilon_2^2\widehat{\mathcal{L}}_{-3}^{\textrm{NS}}\partial_{\theta}
-6\epsilon_2^2\widehat{\mathcal{G}}_{-3/2}^{\textrm{NS}}\partial_x^2
-8\epsilon_2^2\widehat{\mathcal{G}}_{-5/2}^{\textrm{NS}}\partial_x
-6\epsilon_2^2\widehat{\mathcal{G}}_{-7/2}^{\textrm{NS}} +
\\
&\quad
-18\widehat{\mathcal{G}}_{-3/2}^{\textrm{NS}}\widehat{\mathcal{L}}_{-2}^{\textrm{NS}}
+\theta\Big(\epsilon_2^4\partial_x^4
-4\epsilon_2^2\widehat{\mathcal{G}}_{-5/2}^{\textrm{NS}}\partial_x\partial_{\theta}
-4\epsilon_2^2\widehat{\mathcal{G}}_{-7/2}^{\textrm{NS}}\partial_{\theta}
+20\epsilon_2^2\widehat{\mathcal{L}}_{-2}^{\textrm{NS}}\partial_x^2 +
\\
&\quad
+20\epsilon_2^2\widehat{\mathcal{L}}_{-3}^{\textrm{NS}}\partial_x
+12\epsilon_2^2\widehat{\mathcal{L}}_{-4}^{\textrm{NS}}
-9\widehat{\mathcal{G}}_{-3/2}^{\textrm{NS}}\widehat{\mathcal{G}}_{-5/2}^{\textrm{NS}}
+36\big(\widehat{\mathcal{L}}_{-2}^{\textrm{NS}}\big)^2\Big),
\\
\widehat{\mathcal{A}}_{9/2}^{\textrm{NS}}&=
\epsilon_2^5\partial_x^4\partial_{\theta}
+20\epsilon_2^3\widehat{\mathcal{L}}_{-2}^{\textrm{NS}}\partial_x^2\partial_{\theta}
+20\epsilon_2^3\widehat{\mathcal{L}}_{-3}^{\textrm{NS}}\partial_x\partial_{\theta}
+12\epsilon_2^3\widehat{\mathcal{L}}_{-4}^{\textrm{NS}}\partial_{\theta}
-10\epsilon_2^3\widehat{\mathcal{G}}_{-3/2}^{\textrm{NS}}\partial_x^3 +
\\
&\quad
-20\epsilon_2^3\widehat{\mathcal{G}}_{-5/2}^{\textrm{NS}}\partial_x^2
-30\epsilon_2^3\widehat{\mathcal{G}}_{-7/2}^{\textrm{NS}}\partial_x
-24\epsilon_2^3\widehat{\mathcal{G}}_{-9/2}^{\textrm{NS}}
-\epsilon_2\widehat{\mathcal{G}}_{-3/2}^{\textrm{NS}}\widehat{\mathcal{G}}_{-5/2}^{\textrm{NS}}\partial_{\theta} +
\\
&\quad
+36\epsilon_2\big(\widehat{\mathcal{L}}_{-2}^{\textrm{NS}}\big)^2\partial_{\theta}
-110\epsilon_2\widehat{\mathcal{G}}_{-3/2}^{\textrm{NS}}\widehat{\mathcal{L}}_{-2}^{\textrm{NS}}\partial_x
-56\epsilon_2\widehat{\mathcal{G}}_{-3/2}^{\textrm{NS}}\widehat{\mathcal{L}}_{-3}^{\textrm{NS}}
-72\epsilon_2\widehat{\mathcal{G}}_{-5/2}^{\textrm{NS}}\widehat{\mathcal{L}}_{-2}^{\textrm{NS}} +
\\
&\quad
+\theta\Big(\epsilon_2^5\partial_x^5
-10\epsilon_2^3\widehat{\mathcal{G}}_{-5/2}^{\textrm{NS}}\partial_x^2\partial_{\theta}
-20\epsilon_2^3\widehat{\mathcal{G}}_{-7/2}^{\textrm{NS}}\partial_x\partial_{\theta}
-18\epsilon_2^3\widehat{\mathcal{G}}_{-9/2}^{\textrm{NS}}\partial_{\theta}
+40\epsilon_2^3\widehat{\mathcal{L}}_{-2}^{\textrm{NS}}\partial_x^3 +
\\
&\quad
+60\epsilon_2^3\widehat{\mathcal{L}}_{-3}^{\textrm{NS}}\partial_x^2
+72\epsilon_2^3\widehat{\mathcal{L}}_{-4}^{\textrm{NS}}\partial_x
+48\epsilon_2^3\widehat{\mathcal{L}}_{-5}^{\textrm{NS}}
-34\epsilon_2\widehat{\mathcal{G}}_{-5/2}^{\textrm{NS}}\widehat{\mathcal{L}}_{-2}^{\textrm{NS}}\partial_{\theta}
-2\epsilon_2\widehat{\mathcal{G}}_{-3/2}^{\textrm{NS}}\widehat{\mathcal{L}}_{-3}^{\textrm{NS}}\partial_{\theta} +
\\
&\quad
-56\epsilon_2\widehat{\mathcal{G}}_{-3/2}^{\textrm{NS}}\widehat{\mathcal{G}}_{-5/2}^{\textrm{NS}}\partial_x
-56\epsilon_2\widehat{\mathcal{G}}_{-3/2}^{\textrm{NS}}\widehat{\mathcal{G}}_{-7/2}^{\textrm{NS}}
+256\epsilon_2\big(\widehat{\mathcal{L}}_{-2}^{\textrm{NS}}\big)^2\partial_x
+256\epsilon_2\widehat{\mathcal{L}}_{-2}^{\textrm{NS}}\widehat{\mathcal{L}}_{-3}^{\textrm{NS}}\Big).
\label{high_ns_q_ex}
\end{split}
\end{align}
The form of these operators coincides with that of operators that implement classical equations of motion for certain fields in the classical super-Liouville theory \cite{Belavin:2006pv}, which is a nice check of our formalism.


\subsection{Examples of super-quantum curves}

It is also interesting to consider specific potentials in super-eigenvalue models and corresponding super-quantum curves. We consider analogous examples as in the bosonic case in section \ref{ssec-examples}, i.e. super-Gaussian and super-multi-Penner models. More details, in particular the analysis of planar one-cut solutions in these examples, are given in \cite{Ciosmak:2016wpx}.

To start with we consider a supersymmetric version of the Gaussian model, whose potential has a quadratic bosonic term, and includes bosonic and fermionic linear terms depending on bosonic and fermionic times $t$ and $\xi$
\begin{equation}
V_t(x,\theta)=V_{B,t}(x)+V_F(x)\theta,\qquad
V_{B,t}(x)=tx+\frac{1}{2}x^2,\qquad
V_F(x)=\xi x.
\label{s_d_gauss}
\end{equation}
Specializing the super-quantum curve equations (\ref{q_curve_eq_3_2}) at level 3/2 to this case we find
\begin{equation}
\widehat{A}_{3/2}\widehat{\chi}_{\alpha_{\pm}}(x,\theta)=0,\qquad
\widehat{A}_{3/2}=
-\partial_x\partial_{\theta}-\beta^{\pm 1}\widehat{G}_{-3/2}
-\theta\Big(\partial_x^2-2\beta^{\pm 1}\widehat{L}_{-2}\Big),
\label{g_s_curve}
\end{equation}
where $\alpha_{\pm} = \pm\beta^{\pm 1/2}\hbar$, and after some algebra we find
\begin{align}
\begin{split}
\hbar^2\widehat{G}_{-3/2}\widehat{\chi}_{\alpha}(x,\theta)&=
\Big(\xi(x+t)x+Q\hbar\xi-2\xi(\mu-\alpha)
-\frac{\hbar^2}{t}\big(\partial_{\xi}+\partial_{\theta}+(\xi-\theta)\partial_x\big)\Big)\widehat{\chi}_{\alpha}(x,\theta),\\
\hbar^2\widehat{L}_{-2}&=\frac{1}{2}(x+t)^2+\frac{1}{2}Q\hbar-\mu+\alpha.
\label{g_op_gl}
\end{split}
\end{align}

The second example we consider is the super-multi-Penner super-eigenvalue model with the potential
\begin{align}
\begin{split}
V(x,\theta) & =\sum_{i=1}^M \alpha_i \log (x-x_i+\theta_i\theta)
=V_B(x)+V_F(x)\theta,    \\
& V_B(x)=\sum_{i=1}^M\alpha_i\log(x-x_i),\quad
V_F(x)=\sum_{i=1}^M\frac{\alpha_i\theta_i}{x-x_i}.
\label{s_m_penner}
\end{split}
\end{align}
It is convenient to rescale the wave-function in (\ref{chi_hat_def}) and introduce the following normalization
\begin{equation}
\widetilde{\chi}_{\alpha}(x,\theta)=\widehat{\chi}_{\alpha}(x,\theta)
\prod_{i\neq j}(x_i-x_j-\theta_i\theta_j)^{\frac{\alpha_i\alpha_j}{2\hbar^2}}.
\label{chi_til_def}
\end{equation}
Similarly as in the bosonic case, in the super-multi-Penner model the potential term
\be
e^{-\frac{\sqrt{\beta}}{\hbar}\sum_{a=1}^NV(z_a,\vartheta_a)}
\ee
takes an analogous form as $M$ determinant-like insertions of primary fields (\ref{chi_hat_def}). It follows that the wave-function (\ref{chi_til_def}) represents a correlation function in the super-Liouville theory, which involves $(M+2)$ Neveu-Schwarz primary fields inserted on ${\IP}^1$: $M$ of those fields are encoded in the potential, one field is represented by the determinant-like insertion, and one additional field is inserted at $x=\infty \in {\IP}^1$. The primary field at $x=\infty$ can also be removed by the momentum conservation condition, and then the model with the above potential describes super-Liouville theory with $(M+1)$ primary fields. It follows that various objects familiar in super-Liouville theory arise upon the specialization of our formalism to the super-multi-Penner potential. To construct super-quantum curves for the model with the potential (\ref{s_m_penner}), we note first that fermionic $\widehat{h}(x)$ in (\ref{h_x_op}) and the bosonic $\widehat{f}(x)$ in (\ref{f_x_op}) operators take form
\begin{align}
\begin{split}
\widehat{h}(x) &=
\hbar^2\sum_{i=1}^M\frac{1}{x-x_i}D_i,\\
\widehat{f}(x) &=
\hbar^2\sum_{i=1}^M\Big(\frac{1}{x-x_i}\partial_{x_i}
+\frac{\theta_i}{2(x-x_i)^2}\partial_{\theta_i}\Big),
\end{split}
\end{align}
where $D_i=-\partial_{\theta_i}+\theta_i\partial_{x_i}$. Second, we introduce the super-Virasoro generators represented on the wave-function $\widetilde{\chi}_{\alpha}(x,\theta)$ in (\ref{chi_til_def}); taking advantage of (\ref{h_g_chi_rep}), for $n\geq 2$ they take form
\begin{align}
\begin{split}
\widetilde{G}_{-n+1/2} & =
\Big(\prod_{i\neq j}(x_i-x_j-\theta_i\theta_j)^{\frac{\alpha_i\alpha_j}{2\hbar^2}}\Big)
\widehat{G}_{-n+1/2}
\Big(\prod_{i\neq j}(x_i-x_j-\theta_i\theta_j)^{-\frac{\alpha_i\alpha_j}{2\hbar^2}}\Big) =\\
& = \sum_{i=1}^M\bigg(\frac{2(n-1)\Delta_{\alpha_i}\theta_i}{(x_i-x)^n}
-\frac{1}{(x_i-x)^{n-1}}D_i\bigg),   \label{G-Penner}
\end{split}
\end{align}
and
\begin{align}
\begin{split}
\widetilde{L}_{-n} & =
\Big(\prod_{i\neq j}(x_i-x_j-\theta_i\theta_j)^{\frac{\alpha_i\alpha_j}{2\hbar^2}}\Big)
\widehat{L}_{-n}
\Big(\prod_{i\neq j}(x_i-x_j-\theta_i\theta_j)^{-\frac{\alpha_i\alpha_j}{2\hbar^2}}\Big) = \\
& = \sum_{i=1}^M\bigg(\frac{(n-1)\Delta_{\alpha_i}}{(x_i-x)^n}
-\frac{1}{(x_i-x)^{n-1}}\partial_{x_i}
+\frac{(n-1)\theta_i}{2(x_i-x)^n}\partial_{\theta_i}\bigg).   \label{L-Penner}
\end{split}
\end{align}
Super-quantum curves for the super-multi-Penner model can now be constructed using the above representation of super-Virasoro generators in expressions for super-Virasoro singular vectors. For example,  super-quantum curve equations (\ref{q_curve_eq_3_2}) at level $3/2$ take form
\begin{equation}
\widetilde{A}_{3/2}\widetilde{\chi}_{\alpha = \pm \beta^{\pm 1/2}\hbar}(x,\theta)=0,\qquad
\widetilde{A}_{3/2}=
-\partial_x\partial_{\theta}-\beta^{\pm 1}\widetilde{G}_{-3/2}
-\theta\Big(\partial_x^2-2\beta^{\pm 1}\widetilde{L}_{-2}\Big).
\end{equation}
Moreover, using the $\mathfrak{osp}(1|2)$ invariance of the super-eigenvalue model, it is possible to remove some time-dependence from operators $\widetilde{G}_{-n+1/2}$ and $\widetilde{L}_{-n}$. In particular, for $M=2$ and imposing momentum conservation condition, time derivatives can be completely removed, and the action of the above operators on $\widetilde{\chi}_{\alpha}(x,\theta)$ takes for example the following form
\begin{equation}
\widetilde{G}_{-3/2}\widetilde{\chi}_{\alpha}(x,\theta)=
\bigg[-\sum_{i=1,2}\frac{1}{x-x_i}D+\sum_{i=1,2}\frac{2\Delta_{\alpha_i}\theta_i}{(x-x_i)^2}-2\frac{\Delta_{\alpha}\theta+\Delta_{\alpha_1}\theta_1+\Delta_{\alpha_2}\theta_2}{(x-x_1)(x-x_2)}\bigg]\widetilde{\chi}_{\alpha}(x,\theta),
\end{equation}
and
\begin{align}
\widetilde{L}_{-2}\widetilde{\chi}_{\alpha}(x,\theta)&=
\bigg[-\sum_{i=1,2}\frac{1}{x-x_i}\partial_x+\sum_{i=1,2}\frac{\theta_iD+2\Delta_{\alpha_i}}{2(x-x_i)^2}
-\frac{\theta\partial_{\theta}+2\Delta_{\alpha}+2\Delta_{\alpha_1}+2\Delta_{\alpha_2}}{2(x-x_1)(x-x_2)} + \\
&
+\frac{1}{(x-x_1)(x-x_2)}\sum_{i=1,2}\frac{\Delta_{\alpha}\theta_i\theta}{x-x_i}
-\frac{\big((x-x_1)\Delta_{\alpha_1}-(x-x_2)\Delta_{\alpha_2}\big)\theta_1\theta_2}{(x-x_1)^2(x-x_2)^2}\bigg]\widetilde{\chi}_{\alpha}(x,\theta).  \nonumber
\end{align}
Therefore super-quantum curve equations in this case are time-independent, and take form of ordinary super-differential equations. These equations are directly related to the ordinary differential equation considered by Dotsenko and Fateev
\cite{Dotsenko:1984nm,Dotsenko:1984ad}. This equation was analyzed in  \cite{Belavin:2006zr} and some properties of its solutions,
which can be expressed in term of certain two-fold contour integrals, were discussed in \cite{Belavin:2007gz}.

We can also consider the classical limit (\ref{par_h_classic}) in the above example. The super-quantum curves at level $3/2$ reduces in this limit to a system of equations
\begin{align}
\begin{split}
y_B(x)y_F(x) & = \sum_{i=1,2}\frac{\alpha_i^2\theta_i}{(x-x_i)^2}
-\frac{\alpha_1^2\theta_1+\alpha_2^2\theta_2}{(x-x_1)(x-x_2)},    \\
y_B(x)^2+y_F'(x)y_F(x) & = \sum_{i=1,2}\frac{\alpha_i^2}{(x-x_i)^2}
-\frac{\alpha_1^2+\alpha_2^2}{(x-x_1)(x-x_2)} + \\
&\qquad
-\frac{\big((x-x_1)\alpha_1^2-(x-x_2)\alpha_2^2\big)\theta_1\theta_2}{(x-x_1)^2(x-x_2)^2}.
\label{s_curve_m2_l}
\end{split}
\end{align}
The same result can be also found from the analysis of the planar solution of this model \cite{Ciosmak:2016wpx}.


\subsection{Asymptotic expansion}

In the super-eigenvalue model one can introduce the following connected differentials
\begin{align}
W_{(h,0)}(x_1,\ldots,x_h)&=\beta^{h/2}\bigg<\prod_{i=1}^h\sum_{a=1}^N\frac{dx_i}{x_i-z_a}\bigg>^{(\mathrm{c})},
\label{h_conn_diff}
\\
W_{(h,1)}(x_1,\ldots,x_h|x,\theta)&=\beta^{(h+1)/2}\bigg<\prod_{i=1}^h\sum_{a=1}^N\frac{dx_i}{x_i-z_a}\cdot \sum_{a=1}^N\frac{\vartheta_a\theta}{x-z_a}\bigg>^{(\mathrm{c})},
\label{hf_conn_diff}
\end{align}
where $\left<\cdots \right>^{(\mathrm{c})}$ denotes the connected part of the normalized expectation value $\left<\cdots \right>/Z$. From the definition (\ref{chi_hat_def}) we find that (\ref{Psi-def}) has the following expansion
\begin{align}
\log \Psi_{\alpha}(x,\theta)&=\frac{\alpha}{\hbar^2}V(x,\theta) +
\label{recon_wave}\\ 
&\quad
+\sum_{h=1}^{\infty}\frac{1}{h!}\Big(-\frac{\alpha}{\hbar}\Big)^h
\int^{x}_{\infty}\cdots \int^{x}_{\infty}\Big(W_{(h,0)}(x_1,\ldots,x_h)+W_{(h,1)}(x_1,\ldots,x_h|x,\theta)\Big),   \nonumber
\end{align}
analogously to the asymptotic expansion in the bosonic case presented in section \ref{ssec-toprec}.

Note that in the bosonic case the differentials (\ref{h_conn_diff-Vir}) analogous to (\ref{h_conn_diff}) can be reconstructed using the topological recursion, which then gives rise to the quantization procedure sketched in section \ref{ssec-toprec}. A supersymmetric version of the topological recursion that would compute differentials (\ref{h_conn_diff}) and (\ref{hf_conn_diff}) in super-eigenvalue models is not known to date. However, once such a formulation would be established, the expression (\ref{recon_wave}) would enable the construction of super-quantum curves beyond the realm of super-eigenvalue models.


\section{Summary}   \label{sec-summary}

In this work we have shown that quantum curves have the structure of singular vectors of appropriate symmetry algebra. We discussed in detail the case of Virasoro algebra and the Neveu-Schwarz sector of the super-Virasoro algebras. From our perspective quantum curves typically considered in the literature to date correspond simply to level 2 of the Virasoro algebra. In the case of Virasoro algebra we also presented the reformulation of the quantization procedure in terms of the topological recursion.

There are many interesting directions in which these results could be generalized. In \cite{CHMS} we present more explicit formulation of quantum curves from the perspective of conformal field theory, which in particular enables to determine super-quantum curves corresponding to the Ramond sector. Furthermore, considering more general matrix models with more general underlying algebras (e.g. W-algebras) should lead to more general classes of quantum curves.

An interesting outcome of our work is identification of universal $\alpha$-dependent expressions, which reduce to singular vectors for values of $\alpha$ corresponding to degenerate momenta. It is an important problem to simplify the algorithm that enables us to determine these expressions, and generalize them to arbitrary level.

Another important challenge is to find the topological recursion for the super-eigenvalue model. Such a recursion would enable quantization of a general class of supersymmetric algebraic curves, not necessarily associated to super-eigenvalue models.

It is also important to understand in more detail relations of our formalism to surface operators and brane systems, sketched briefly in the introduction.

We are convinced that higher level quantum curves presented in this work should play important role also in enumerative geometry, integrable systems, knot theory, and many other systems where the topological recursion has already proved very useful.


\bigskip

{\bf Acknowledgements.} We are grateful to the American Mathematical Society for the organization of the 2016 AMS von Neumann Symposium \emph{``Topological Recursion and its Influence in Analysis, Geometry, and Topology''}. P.S. is indebted to Bertrand Eynard, Chiu-Chu Melissa Liu, and Motohico Mulase for providing the opportunity to present these results, inspiration and encouragement. We also thank Vincent Bouchard, Zbigniew Jask{\'o}lski and Chaiho Rim for discussions. This work is supported by the ERC Starting Grant no. 335739 \emph{``Quantum fields and knot homologies''} funded by the European Research Council under the European Union's Seventh Framework Programme.


\bibliographystyle{JHEP}
\bibliography{abmodel}

\end{document}